\documentclass[traditabstract]{aa}
\usepackage{graphicx}
\usepackage{txfonts}
\usepackage{subfigure}

\begin{document}

   \title{SiO excitation from dense shocks in the  earliest stages of massive star formation\thanks{Based on observations made with ESO telescopes at the La Silla Paranal
Observatory under programme ID 089.C-0203}}

   \author{S. Leurini
          \inst{1}
          \and
          C. Codella\inst{2} \and A. L\'opez-Sepulcre\inst{3,4} \and A. Gusdorf\inst{5} \and T. Csengeri\inst{1} \and S. Anderl\inst{3,4}}

   \institute{Max-Planck-Institut f\"ur Radioastronomie, Auf dem H\"ugel 69, 53121 Bonn, Germany\\\email{sleurini@mpifr.de}
    \and INAF, Osservatorio Astrofisico di Arcetri, Largo E. Fermi 5, 50125, Firenze, Italy
    \and Univ. Grenoble Alpes, IPAG, F-38000 Grenoble, France
\and CNRS, IPAG, F-38000 Grenoble, France
    \and LERMA, UMR 8112 du CNRS, Observatoire de Paris, \'Ecole Normale Sup\'erieure, 24 rue Lhomond, 75231, Paris Cedex 05, France
}
   \date{\today}

  \abstract{Molecular outflows are a direct consequence
    of accretion, and therefore they represent one of the best tracers of accretion processes 
in the still poorly understood early phases of high-mass star formation.   
    Previous studies suggested that the SiO
    abundance decreases with the evolution of a massive young stellar
    object probably because of a decay of jet activity, as witnessed in low-mass star-forming regions. 
    We investigate the SiO excitation conditions and its
    abundance in outflows from a sample of massive young stellar objects 
through observations of the SiO(8--7) and
    CO(4--3) lines  with the
    APEX telescope. Through a non-LTE analysis, we find that the excitation conditions of SiO increase with the velocity of the emitting gas.
 We also compute the SiO abundance through the SiO and CO integrated intensities at high velocities. For the sources in our sample we find no
 significant variation  of the SiO abundance with evolution for a bolometric luminosity-to-mass ratio of between 4 and 50\,$L_\odot/M_\odot$. We also find a weak increase of the SiO(8--7) luminosity with the bolometric luminosity-to-mass ratio. We speculate that this might be explained with an increase of density in the gas traced by SiO.  We find that the densities constrained by the SiO observations require the use of shock models that include grain-grain processing. For the first time, such models are compared  and found to be compatible with SiO observations. A pre-shock density of $10^5\, $cm$^{-3}$ is globally inferred from these comparisons. Shocks with a velocity higher than 25~km~s$^{-1}$ are invoked for the objects in our sample where the SiO is observed with a corresponding velocity dispersion. Our comparison of shock models with observations suggests that  sputtering of silicon-bearing material (corresponding to less than 10\% of the total silicon abundance) from the grain mantles is occurring.}

   \keywords{stars: formation --
                ISM: jets and outflows --
                ISM: molecules}

   \maketitle
%

\section{Introduction}\label{intro}

The formation mechanism of high-mass stars ($M > 8$\,M$_\odot$) has
been an open question for the past few decades, the main reason being
that the strong radiation pressure exerted by the young massive star
overcomes the star's gravitational attraction \citep{1974A&A....37..149K}.
Different mechanisms have been proposed to lead to the formation of
massive stars, some based on disc-mediated accretion
\citep[e.g.,][]{2003ApJ...585..850M,2005Natur.438..332K}, others on
alternative scenarios where massive cores accrete mass from their
surroundings while collapsing \citep[e.g.,][]{2006MNRAS.370..488B}.

Molecular outflows are predicted in both scenarios and can help 
 distinguish between
models of massive star-formation.  
 In particular,
a study of their properties and their
evolution with  time  would allow us to understand the different phases
involved in the formation of massive stars. A full understanding of the properties of outflows as function of mass of the powering source would also help to evaluate whether
there are similarities with low-mass star formation because the relations derived for low-mass young stellar objects (YSOs) \citep[e.g.,][]{1996A&A...311..858B} should hold over the whole range of masses if massive stars form in the same way as low-mass stars. Given their relatively large scales ($\sim
1$\,pc), bipolar molecular outflows often represent the easiest, if not
the only, means to observationally investigate high-mass star
formation in a statistically significant number of objects.

SiO thermal emission is the best tool to investigate molecular outflows. Its appearance in the gas phase is attributed
to sputtering of silicon-bearing material from grains \citep{Gusdorf081,Gusdorf082}, supplemented by grain-grain interactions \citep[e.g. vaporisation and shattering,][]{Guillet07,Guillet09,Guillet11,Anderl:2013} in higher-density regions.
 As a result, SiO
suffers minimal contamination from the more quiescent gas. 
Observations towards low-mass protostars point to an evolutionary
picture in which bipolar jets and outflows are always associated with
young Class 0 objects \citep[e.g.,
][]{1999A&A...343..571G,2007A&A...462L..53C}, while Class I/II objects
show less evidence of powerful jets. Finding an
analogous trend in high-mass YSOs would be
strongly suggestive of a similar formation scenario for low- and
high-mass stars.

In an attempt to investigate how the properties of molecular outflows
 change with the evolution of the source in massive YSOs,
\citet{2011A&A...526L...2L} and \citet{2013A&A...557A..94S} 
investigated massive parsec-scale molecular clumps in SiO transitions with the
IRAM 30\,m telescope. 
They concluded that the SiO abundance decreases with
time, favouring a scenario in which the
amount of SiO is mainly enhanced in the first evolutionary stages,
probably by strong shocks produced by the protostellar jet, and
decreases as the object evolves, most likely because of a decaying jet
activity.

A limitation of these  studies is that they used low -excitation SiO lines with upper level energies ranging from 6\,K ($J=2-1$)\footnote{All molecular parameters presented in this paper are from the JPL catalogue \citep{pickett_JMolSpectrosc_60_883_1998}} to 30\,K ($J=5-4$). 
Models of SiO emission show that a more
exhaustive study of the excitation of SiO needs higher $J$ transitions:
\citet{2007A&A...462..163N} demonstrated that the SiO(8--7) and
SiO(3--2) lines can be used to distinguish between different
excitations in shocks as they cover a wider energy range 
($E_{\rm{low}_{\rm{SiO(3-2)}}}\sim$12\,K, $E_{\rm{low}_{\rm{SiO(8-7)}}}\sim$73\,K). Higher $J$ SiO lines are also needed to
distinguish between different shock conditions \citep{Gusdorf081}. 
In the current study, we present observations of a sub-sample of
sources from \citet{2011A&A...526L...2L}  in the CO(4--3) and SiO(8--7) transitions with the APEX
telescope with a resolution that well matches that of  the SiO(3--2) data presented by \citet{2011A&A...526L...2L}. 
We aim  to study the SiO(3--2) and SiO(8--7) data {\it(1)} to
verify  whether the decrease in SiO luminosity with evolution seen in previous studies  is confirmed
when 
excitation effects are taken into account, {\it(2)}  to derive a direct measurement of the SiO abundance using CO(4--3),  {\it(3)} to determine whether the SiO abundance changes with evolution, {\it(4)} to discuss the properties of SiO as function of evolution, and {\it(5)} to compare for the first time 
 SiO observations with shock models that include grain-grain processing at the high densities typical of high-mass star-forming clumps.
 In Sect.\,\ref{sample} we discuss our selection criteria and present the properties of 
the observed sources. In Sect.\,\ref{obs} we present the observations, while in
Sect.\,\ref{res} we discuss the observational results. Finally in Sect.\,\ref{dis} we compute the SiO line luminosities and abundances for our sources, study their behaviour as function of evolution, and interprete our observations in the framework of shock models and a non-LTE analysis. 

\begin{figure}
\centering
\includegraphics[width=9cm]{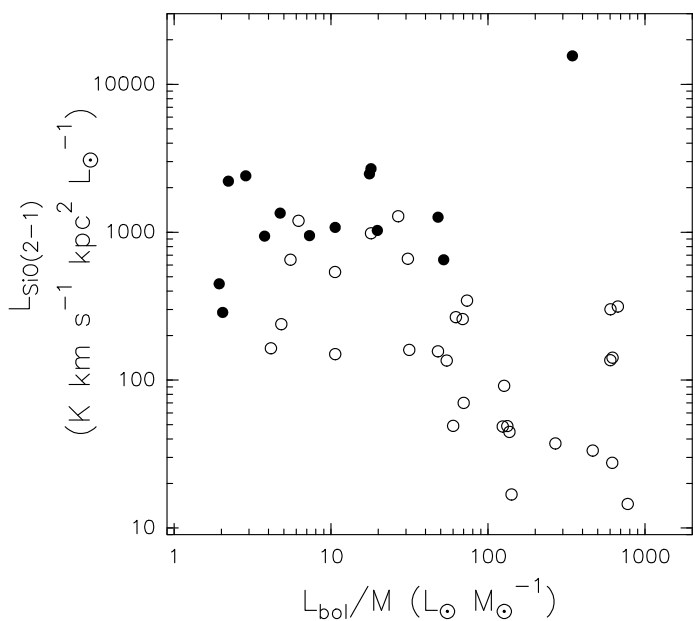}
\caption{SiO(2–1) luminosity against $L_{\rm{bol}}/M$ for the full sample of sources studied by \citet{2011A&A...526L...2L}.
Full circles depict the sources observed in SiO(8--7) that we
studied here.}\label{ana}
\end{figure}

\section{Selection of the sample}\label{sample}
From the sample of \citet{2011A&A...526L...2L}, we selected fifteen sources with 
a line-width  in SiO(3--2) broader than 40\,km\,s$^{-1}$ and
a peak intensity 
higher than 0.2\,K. The third criterion was that the sources were
observable with the APEX telescope\footnote{APEX is a collaboration between
the Max-Planck-Institut f\"ur Radioastronomie, the European Southern
Observatory, and the Onsala Space Observatory}.  The list of observed sources and their properties is given in Table\,\ref{sources}.  Luminosities and masses come from \citet{2011A&A...526L...2L}  and, where available,  from 
\citet{2013A&A...557A..94S}, who improved the previous spectral energy distribution fits with data from the {\it Herschel} infrared Galactic Plane
Survey \citep{2010A&A...518L.100M}. For IRAS\,18151-1208\_2, there were not enough continuum data available to model 
the spectral energy distribution, which resulted in a lack of mass and luminosity estimates for this source.  The source G19.61–0.24A was assumed to be at the near distance of 3.8\,kpc by \citet{2011A&A...526L...2L}. However, \citet{2003ApJ...582..756K} located it at the far distance of 12.6\,kpc thanks to 
HI interferometric observations,  which show absorption features
at velocities higher than the systemic velocity. 
Therefore, we  here
adopt the far distance for the source, and correct the mass and luminosity from \citet{2011A&A...526L...2L}  for the new value. We note, however, that the mass and luminosity of G19.61--0.24A have large uncertanties because of the complexity of the source and the coarse spatial resolution of the data used to fit its spectral energy distribution. For example, \citet{2011A&A...525A..72f} estimated a luminosity of $\sim 10^5$\,L$_\odot$ for the powering source of the main outflow in the G19.61--0.24A cluster, while the total mass of the clumps in the region is 1580\,M$_\odot$ from interferometric measurements.

The variation of the SiO(2--1) luminosity  as function of the  bolometric luminosity-to-mass ratio, $L_{\rm{bol}}/M$ (a rough estimator of evolution in the star formation process, e.g., \citealt{2008A&A...481..345M,2013ApJ...779...79M}), for the selected sources is shown in Fig.\,~\ref{ana}. Although 
our selection criteria bias us towards the more luminous sources in SiO(2--1) of the sample of \citet{2011A&A...526L...2L}, we cover a wide range of 
 $L_{\rm{bol}}/M$ values up to $\sim 350$\,$L_\odot/M_\odot$.

To probe the evolutionary phase of the clumps, we investigated the emission at 22 and 8\,$\mu$m in the 
 WISE and Midcourse Space eXperiment (MSX)  missions. We expect the youngest sources in the sample to be  still dark at 22\,$\mu$m and the most evolved to be detected at 22 and 8\,$\mu$m.
The association of the sources
with mid-IR emission at 21--22\,$\mu$m is described by \citet{2014A&A...565A..75C}. Sources are classified as 22\,$\mu$m-quiet if there is no association with a WISE source and
as 22\,$\mu$m-loud in case of association, and correspondingly 
infrared loud at 8\,$\mu$m (8\,$\mu$m-loud) and 
infrared quiet at 8\,$\mu$m (8\,$\mu$m-quiet) if associated with 8\,$\mu$m emission or not.
  Table\,\ref{sources} summarises the IR properties of the sources.

\begin{table*}
\caption{Properties of the observed sources.}\label{sources}      
\centering
\begin{tabular}{lrrrrrrcc}
\hline \hline
Source & R.A. [J2000]& Dec. [J2000] & v$_{LSR}$ &d$^{\rm{a}}$ & M$^{\rm{b}}$& L$_{\rm{bol}}^{\rm{b}}$&IR-$8\mu$m&IR-$22\mu$m\\
&&&(km\,s$^{-1}$)&(kpc)&(M$_\odot$)&(L$_\odot$)&\\
\hline
G19.27+0.1M2        &18:25:52.60 &--12:04:48.0& 26.9 &2.4          &  60&   200                   &quiet&quiet\\ %
G19.27+0.1M1        &18:25:58.50 &--12:03:59.0& 26.5 &2.4          & 200&   400                   &quiet&quiet\\ %
G19.61--0.24A        &18:27:38.16 &--11:56:40.2& 42.4 &12.6$^{\rm{c}}$ & 10000$^{\rm{d}}$&3\,400\,000$^{\rm{d}}$&loud&loud\\ %
G20.08--0.14         &18:28:10:28 &--11:28:48.7& 42.5 &3.4$^{\rm{e}}$ & 400$^{\rm{a}}$ & 19\,700$^{\rm{a}}$&loud&loud\\ %
IRAS\,18264--1152    &18:29:14.40 &--11:50:21.3& 43.9 &3.5          &1600           & 19\,000&loud&loud\\ %
G23.60+0.0M1        &18:34:11.60 &--08:19:06.0&106.5 &6.2$^{\rm{f}}$&1800& 5200&quiet&quiet\\%
IRAS\,18316--0602    &18:34:20.46 &--05:59:30.4& 42.5 &3.1          &1600&32\,000&loud&loud\\ %
G23.60+0.0M2        &18:34:21.10 &--08:18:07.0& 53.6 &3.9          & 300& 3000&quiet&quiet\\ %
G24.33+0.1M1        &18:35:07.90 &--07:35:04.0&113.6 &6.7$^{\rm{f}}$ &2600&48\,000&quiet&loud\\ %
G25.04--0.2M1        &18:38:10.20 &--07:02:34.0& 63.8 &4.3$^{\rm{f}}$ & 400$^{\rm{a}}$& 1300$^{\rm{a}}$&quiet&quiet\\ %
G34.43+0.2M1        &18:53:18.00 &+01:25:23.0& 58.1 &3.7          &1400&24\,000&quiet&loud\\ %
IRAS\,18507+0121    &18:53:19.58 &+01:24:37.1& 58.2 &3.7          &3100&14\,500&loud&loud\\ 
G34.43+0.2M3        &18:53:20.40 &+01:28:23.0& 59.4 &3.7          & 600&1400&quiet&quiet\\ %
IRAS\,19095+0930    &19:11:54.02 &+09:35:52.0& 44.0 &3.3          & 1000&50\,600&loud&loud\\ %
IRAS\,18151--1208\_2 &18:17:50.50 &--12:07:55.0& 29.8 &3.0          & $-$&$-$&quiet&--\\ %

\hline
\end{tabular}
\tablefoot{
\tablefoottext{a}{\citet{2011A&A...526L...2L}}
\tablefoottext{b}{\citet{2013A&A...557A..94S}}
\tablefoottext{c}{\citet{2003ApJ...582..756K}}
\tablefoottext{d}{mass and bolometric luminosity from \citet{2011A&A...526L...2L}  for 3.8\,kpc are rescaled to 12.6\,kpc}
\tablefoottext{e}{\citet{1996A&AS..120..283H}}
\tablefoottext{f}{\citet{2010A&A...517A..66L}}
}

\end{table*}

\section{Observations}\label{obs}
All observations were performed with the APEX telescope. The majority of the data  was observed under Max Planck Society (MPS) time in April and November 2011  in wobbler-switching
mode, with a wobbler throw of 120\arcsec.  Additional data were taken under ESO programme 089.C-0203
in September 2012.  For the observations performed in April and
November 2011, the FLASH$^+$ receiver was used \citep{klein}. This is a dual-frequency
MPIfR principal investigator receiver that operates simultaneously
in the 345\,GHz and the 460\,GHz atmospheric windows. The 345\,GHz
channel is a sideband-separating detector with 4\,GHz bandwidth per
sideband.  The 345\,GHz channel was tuned to 347.33\,GHz in upper sideband to observe SiO(8--7) and CO(3--2),  and C$^{17}$O(3--2) in lower side band.   The 460\,GHz channel was only available
for the observing run in October 2011 and it was tuned to 461.04\,GHz in lower
sideband to observe CO(4--3).  G19.27+0.1M1,
IRAS\,18316–0602, and G34.43+0.2M3 were not observed in CO(4--3).The ESO observations in SiO(8--7) and CO(3--2) were performed with the APEX-2 receiver \citep{2008A&A...490.1157V}.

\begin{table}
\caption[]{Summary of the observations.}\label{lines}
\begin{tabular}{lccccl}
\hline
\multicolumn{1}{c}{Transition} &
\multicolumn{1}{c}{$\nu$$^{\rm a}$} &
\multicolumn{1}{c}{$E_{\rm u}$$^a$} &
\multicolumn{1}{c}{Beam}&
\multicolumn{1}{c}{$\delta \varv$}&
\multicolumn{1}{c}{r.m.s.}\\ 

\multicolumn{1}{c}{ } &
\multicolumn{1}{c}{(MHz)} &
\multicolumn{1}{c}{(K)} &
\multicolumn{1}{c}{(\arcsec)} &
\multicolumn{1}{c}{(km\,s$^{-1}$)}&
\multicolumn{1}{c}{(K)}\\ 
\hline
CO(3--2) & 345795.99 & 33 &18.0  &2&0.01\\ 
C$^{17}$O(3--2) & 337061.13 & 32 &18.5 &2&0.01 \\ 
CO(4--3) & 461040.77 & 55 &13.5  &1&0.1\\ 
SiO(8--7) & 347330.63 & 75 &18.0 &2&0.01 \\ 
\hline
\end{tabular}
\tablefoot{
\tablefoottext{a}{Frequencies and energies come from the Jet
Propulsion Laboratory (JPL) molecular database \citep{pickett_JMolSpectrosc_60_883_1998}}}
\end{table}

After comparison of the ESO and MPS data at 345\,GHz source by source
for the objects observed in both programmes, observations were
averaged to increase the signal-to-noise ratio.  Data were
converted into $T_{\rm{MB}}$ units assuming a forward efficiency of 0.95 for
both receivers, and a beam efficiency of 0.75 for SiO(8--7), CO(3--2) and C$^{17}$O(3--2), 
and of 0.60 for CO(4--3). The original velocity resolution of
the data was 0.07\,km\,s$^{-1}$ for SiO(8--7), CO(3--2), and C$^{17}$O(3--2), and 0.1\,km\,s$^{-1}$   for
CO(4--3). The spectra were smoothed to a final velocity
resolution of 2\,km\,s$^{-1}$ in the low-frequency data, and of 1\,km\,s$^{-1}$ for
CO(4--3).  The typical r.m.s. of the data at these spectral
resolutions is 0.01\,K at 335 and 345\,GHz, and 0.1\,K at 460\,GHz. The observations are summarised in Table\,\ref{lines}.  Data reduction and analysis were performed with  the GILDAS software\footnote{http://www.iram.fr/IRAMFR/GILDAS}.

\section{Observational results}\label{res}

\subsection{SiO(8--7)}\label{obs87}

Figures\,\ref{spectra1} and \ref{spectra2} show all SiO(8--7) detected spectra of our sample
together with the SiO(3--2) data of \citet[][beam\footnote{http://www.iram.es/IRAMES/mainWiki/Iram30mEfficiencies}=$19\arcsec$]{2011A&A...526L...2L} and with the SiO(5--4) line from \citet{2013A&A...557A..94S}, when available, smoothed to the SiO(8--7) resolution.
The SiO(8--7) line is
detected above 4\,$\sigma$ towards all sources except three (G25.04–0.2M1, G23.60+0.0M2,
and G19.27+0.1M2, Fig.\,\ref{nondet}). 
These results give us a detection rate of $\sim 80\%$, lower
than the 88\% rate found in \citet{2011A&A...526L...2L} using lower-$J$ SiO emission, but higher than the 60\% detection rate found by \citet{2012A&A...538A.140K} in a sample of IR-loud high-mass YSOs
in SiO(8--7) with a lower sensitivity limit than ours. We note, however, that 
the detection rate for our sample
would decrease to 66\% with the sensitivity of their observations.

The three
undetected sources are classified as dark at  8\,$\mu$m and  22\,$\mu$m, which suggests that they are  in a very early evolutionary phase. The non-detection of SiO(8--7) in these very early evolutionary phases 
seems to contradict the findings of \citet{2011A&A...526L...2L}, who reported 
that the SiO luminosity decreases with the evolution of the source and is highest in the earliest stages of star formation. 
In Sect.\,\ref{dis} we investigate  how the luminosity of the SiO(8--7) 
transition and  the SiO abundance  vary with time to determine the effects of excitation on the results of \citet{2011A&A...526L...2L}  and understand our low detection rate in IR-quiet sources.

\begin{table*}
\caption{SiO(8--7) and (3--2) line parameters.}\label{sio87}      
\centering
\begin{tabular}{lrrrr}
\hline \hline
Source &\multicolumn{2}{c}{SiO(8--7)}&\multicolumn{2}{c}{SiO(3--2)}\\
&$FWZP$&$\int{F_{\rm{v}}d\rm{v}}$&$FWZP$&$\int{F_{\rm{v}}d\rm{v}}$\\
&(km\,s$^{-1}$)&(K\,km\,s$^{-1}$)&(km\,s$^{-1}$)&(K\,km\,s$^{-1}$)\\
\hline

G19.27+0.1M2       &-- &--        &51.0 &3.7\\ 
G19.27+0.1M1       &30 &$0.79$ &69.9 &9.1\\ 
G19.61--0.24A       &57 &$9.4$  &59.0 &17.0\\ 
G20.08--0.14        &44 &$2.4$  &50.0 &16.2\\  
IRAS\,18264--1152   &93&$3.5 $  &40.1 &6.2\\ 
G23.60+0.0M1       &15 &$0.44$ &82.9 &6.1\\ 
IRAS\,18316--0602  &34 &$0.85$  &60.3 &9.3\\
G23.60+0.0M2       &-- &--        &57.9 &4.5\\ 
G24.33+0.1M1      &44&$1.08$         &55.0 &5.4\\ 
G25.04--0.2M1       &-- &--        &52.1 &3.4\\ 
G34.43+0.2M1       &49 &$3.3 $  &63.0 &18.\\ 
IRAS\,18507+0121   &50 &$1.11$  &52.1 &6.4\\ 
G34.43+0.2M3       &56 &$1.77$  &72.0 &13.9\\ 
IRAS\,19095+0930   &48 &$1.55$  &55.8 &7.1\\ 
IRAS\,18151--1208\_2&99&$3.2$  &103.1 &11.6\\

\hline
\end{tabular}
\end{table*}

SiO(8--7) emission is typically very broad ,
with an average full width at
zero power ($FWZP$) of 45\,km\,s$^{-1}$. 
Table\,\ref{sio87} lists the measured SiO(8--7) $FWZP$s (and for comparison the SiO(3--2) parameters as well) and the velocity-integrated intensities over the $FWZP$ velocity range for all detected sources.  We note that the blue-shifted profile of SiO(8--7) in  G24.33+0.1M1 shows a feature at 347.35\,GHz (full width at half maximum of 10\,km\,s$^{-1}$) that is neither detected in SiO(3--2), nor in the (5--4) line (Fig.\,\ref{spectra1}). According to the JPL catalogue, this feature might be acetaldehyde (CH$_3$CHO) ($18_{5,13}-17_{5,12}$) and ($18_{5,14}-17_{5,13}$) at 347.346\,GHz and 347.349\,GHz, respectively, with an upper level energy of about 214\,K. Since
there is no other lower energy CH$_3$CHO line in our spectra we cannot unambiguously identify this feature with CH$_3$CHO. To determine the level of possible contamination from this feature for the whole sample, we compared the integrated intensity over a velocity range of 10\,km\,s$^{-1}$ centred at 347.35\,GHz with the integrated intensity over the whole $FWZP$. Contamination is of the order of 7-8\,\% of the total emission for all sources, except for G24.33+0.1M1 (33\,\%), IRAS\,18316--0602 (30\,\%), G34.43+0.2M1 (21\,\%), and
G19.61--0.24A (20\,\%).

\subsection{CO and C$^{17}$O lines}\label{co}

Figures\,\ref{spectra1} and \ref{spectra2} show the spectra of
the CO isotopologues for the sources 
 where SiO(8--7) was detected. Table\,\ref{cos} lists line-widths of the CO(4--3)
and C$^{17}$O(3--2) lines. CO(3--2) profiles are not analysed here because
they are often severely affected by absorptions caused by the off
position used for the observations. As expected, the CO profiles show
very prominent high-velocity wings in all cases. Extreme sources are 
IRAS\,18264--1152 and IRAS\,18151--1208\_2 with $FWZP$ line-widths in excess of 100\,km\,s$^{-1}$.
On the other hand, the  
C$^{17}$O(3--2) line usually traces gas associated with the dense
envelope. Exceptions are IRAS\,18264--1152, G20.08--0.14, G19.61–0.24A,
and IRAS\,19095+0930, where the C$^{17}$O(3--2) transition also
shows
non-Gaussian wings.

The line profiles of  SiO(8--7) and CO(4--3)
are very similar in particular for IRAS 18264-1152, with emission at high velocities  up to $|\varv-\varv_{\rm  {LSR}}|=$\,63\,km\,s$^{-1}$ in the red-shifted wing.
For  IRAS\,18151--1208\_2, the SiO(8--7) and CO(4--3) transitions have the same behaviour
at high velocities, with emission  up to $|\varv-\varv_{\rm  {LSR}}|=$\,40--45\,km\,s$^{-1}$. 
 Emission from CO and SiO often arises from different components in outflows: the former traces swept-up material, the latter traces molecular gas close to the primary jet \citep{2006ApJ...636L.137P}. However, the difference between the two species
 strongly depends on the considered CO transition and  velocity range.
Recent observations show 
that high-velocity CO emission may coincide with SiO even in the $J=2-1$ line (Fig.\,1 of \citealt{2012A&A...548L...2C}). 
In addition, the probability that CO emission traces the same component as SiO increases with $J$ \citep[e.g.,][]{2006ApJ...639..292L}.  Therefore, in Sect.\,\ref{dis} we assume that the SiO(8--7) and CO(4--3) emissions
at high velocity 
originate from the same gas and derive the abundance of SiO 
by means of the SiO(8--7) to CO(4--3) line ratio at high velocities.

\begin{table}
\caption{Parameters of the CO(4-3) and C$^{17}$O(3-2) transitions.}\label{cos}      
\centering
\begin{tabular}{lrr}
\hline \hline
Source &$FWZP_{\rm{CO(4-3)}}$&$FWZP_{\rm{C^{17}O(3-2)}}$\\
&(km\,s$^{-1}$)&(km\,s$^{-1}$)\\
\hline
G19.27+0.1M2       &30 &6 \\ 
G19.27+0.1M1       &-- &6 \\ 
G19.61--0.24A       &87 &55 \\ 
G20.08--0.14        &35 &25 \\  
IRAS\,18264--1152   &170 &19 \\ 
G23.60+0.0M1$^{\rm{a}}$       &50 &15 \\ 
IRAS\,18316--0602   &-- &17 \\ 
G23.60+0.0M2$^{\rm{b}}$        &&8 \\ 
G25.04--0.2M1       &12 &8 \\ 
G34.43+0.2M1       &42 &17 \\ 
IRAS\,18507+0121   &37 &15 \\ 
G34.43+0.2M3       &-- &6 \\ 
IRAS\,19095+0930   &57 &24 \\ 
IRAS\,18151-1208\_2&132 &9 \\
\hline
\end{tabular}
\tablefoot{
\tablefoottext{a}{only red-shifted emission is considered because blue-shifted emission is most likely contaminated by absorption from off position}\tablefoottext{b}{The CO(4-3) line profile is contaminated by off-position absorption.}}
\end{table}

\section{Analysis}\label{dis}

\subsection{SiO luminosity}\label{time}

Following the same approach as \citet{2011A&A...526L...2L}, we investigated the
dependence of  the SiO(8--7) luminosity
$(L_{\rm{SiO87}}=\int{{\rm{SiO}}(8-7)dv}\times 4 \pi d^2)$ 
on the
evolutionary phase of the source  described by   
$L_{\rm{bol}}/M$.  
Although \citet{2004A&A...426...97F} suggested that this ratio is  mainly determined by the most luminous member of the cluster in high-mass star-forming regions, several authors  proposed $L_{\rm{bol}}/M$
to be a measure of the evolutionary phase in the star formation process \citep[e.g.,][]{2008A&A...481..345M,2013ApJ...779...79M}.
Since the APEX telescope has a beam size of  18\arcsec at the frequency of
SiO(8--7) and the beam size of the IRAM  30\,m telescope is 19\arcsec\ at the
frequency of SiO(3--2), the two datasets
are similarly   affected by beam dilution if the two lines are emitted by the same gas. Therefore, we can   investigate how excitation
evolves with time through the SiO(8--7) to SiO(3--2) integrated intensity ratio without any assumption on the size of the emitting gas. In the following, we use the values reported in Table\,\ref{sio87}, which are integrated over the whole line profile, including the blue-shifted emission. This is because possible contamination is only up to $\sim$20--30\% of the total integrated intensity for four sources, while for the others is negligible. The following results are also valid when only the red-shifted wing is considered,
however.

\begin{figure*}
\centering
\includegraphics[angle=-90,width=12cm]{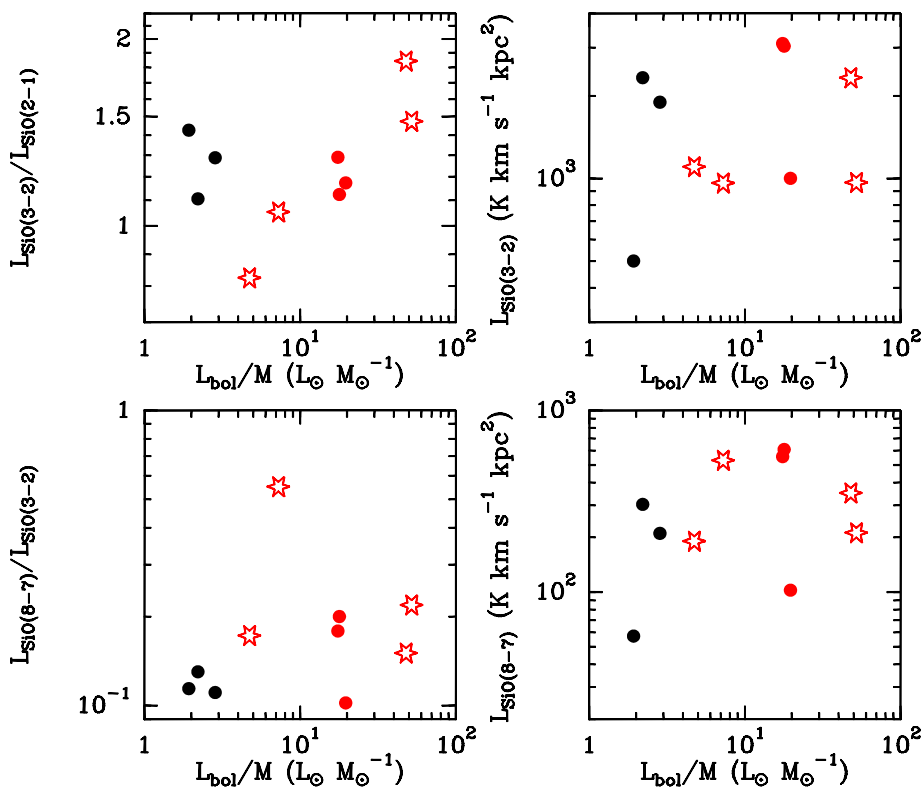}
\caption{Variation of the SiO(8--7) to SiO(3--2) luminosity ratio (bottom left), of the  SiO(8--7) luminosity  (bottom right), of the  SiO(3--2) to SiO(2--1) luminosity ratio (top left), and of the SiO(3--2) luminosity  (top right) as a function of $L_{\rm{bol}}/M$.
 22\,$\mu$-quiet and 22\,$\mu$-loud sources are marked as black and red filled circles. Red stars are 8\,$\mu$-loud sources.
}\label{siolum}
\end{figure*}

Figure\,\ref{siolum} shows  $L_{\rm{SiO87}}$ and the SiO(8--7) to SiO(3--2) integrated intensity ratio as function of $L_{\rm{bol}}/M$.
For comparison, we also show the SiO(3--2) luminosity and the SiO(3--2) to SiO(2--1) 
ratio taken from \citet{2011A&A...526L...2L}  and without any correction for the different beam sizes.  
Given the large uncertanties associated with the mass and the luminosity estimates of G19.61--0.24A (Sect.\,\ref{sample}), we conservatively decided to exclude the source from the analysis.

 In all cases, the SiO luminosities and their ratios seem to weakly increase with L$_{\rm{bol}}/M$. The 
SiO(8--7) luminosity, for example, changes by more than a factor 10 in the sample with an average value of $\sim 300$\,K\,km\,s$^{-1}$\,kpc$^2$. On the other hand, the SiO(3--2) luminosity increases by only a factor 6 (with an average value of 1700\,K\,km\,s$^{-1}$\,kpc$^2$). A similar comparison with the SiO line ratios is complicated by the different dilution factors that affect the SiO(3--2) to SiO(2--1) ratio. Without correcting for the different beams, the SiO(3--2) to SiO(2--1) ratio varies within 
a factor of less than three in our sample of sources. However, we stress that not taking into account the different beam sizes may strongly affect these findings.
On the other hand, the corresponding SiO(8--7) to SiO(3--2)
ratio is not affected by this problem as long as the two lines are emitted by the same gas; this ratio changes by a factor 5 in the sample.

These trends indicate an increase of excitation conditions with $L_{\rm{bol}}/M$. This interpretation is strengthened  by the fact that the only sources not detected in SiO(8--7) are 22\,$\mu$m-dark and therefore probably are in a very early evolutionary phase (see Sect.\,\ref{sio87}).  
The increase of SiO(8--7) luminosity with evolution agrees with the result of \citet{2012A&A...538A.140K}, who found  
that the integrated intensity of the SiO(8--7) line increases with time for a sample of
IR-loud high-mass star-forming regions  (but three
sources, IRAS\,18264--1152, G19.61--0.24A, and G20.08--0.14 are 
common to both papers). Unfortunately, our statistics is too small to conclude about this.

\subsection{SiO abundance}\label{xsio}

\begin{figure}
\centering
\includegraphics[bb = 33 27  536 516,clip,angle=-90,width=8cm]{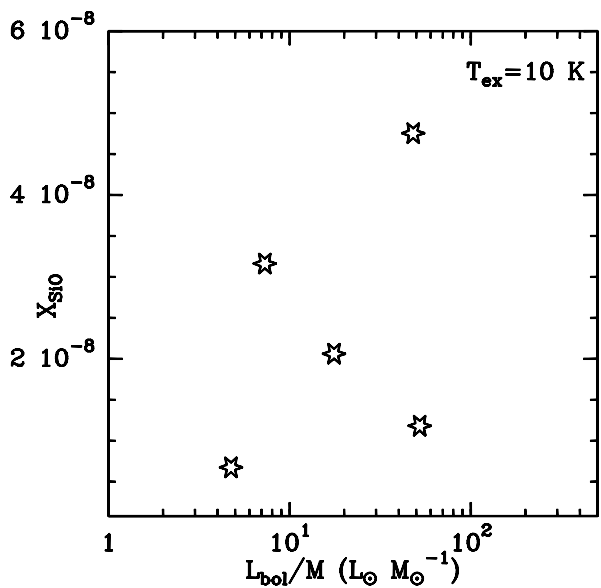}
\caption{Distribution of SiO abundance derived from the ratio of SiO(8--7) to CO(4--3) as function of $L_{\rm{bol}}/M$.}\label{ab}
\end{figure}

Assuming that the SiO(8--7), and CO(4--3) lines are emitted by the same gas at high-velocities (see Sect.\,\ref{co}),   
we computed the SiO abundance relative to CO 
in the wings of the SiO(8--7) and CO(4--3) lines and then converted 
it into abundance relative to H$_2$, $X_{\rm{SiO}}=N_{\rm{SiO}}/N_{\rm{H_2}}$, assuming and abundance of CO to H$_2$ of 10$^{-4}$ \citep{1994ApJ...428L..69L}. 
 To estimate the CO and SiO column densities, we used their integrated  intensities over the same velocity range (Table\,\ref{abu}). 
 Only the red-shifted emission is considered. We excluded G23.60+0.0M1 from the analysis because the SiO(8--7) transition is too narrow compared with CO(4--3).
We calculated the rotation temperature of SiO 
for each source from the intensity ratio of the SiO (8--7) and (3--2) lines
  in the 
velocity range defined on the CO(4--3) line. We checked results in
the optically thin and thick limit and found that apart for
two sources (G19.61--0.24A and IRAS\,18264--1152) the estimated
temperature does not significantly depend on the optical
depth. Typical rotation temperatures, $T_{\rm{rot}}$, are approximately 5\,K, and 11\,K
for G19.61--0.24A and IRAS\,18264--1152 under the assumption of
optically thin emission. For the same sources,
\citet{2013A&A...557A..94S} found an average temperature of 10\,K based on lower $J$ lines.
For the estimate of $X_{\rm{SiO}}$, we assumed a rotation temperature of 10\,K for SiO and CO. Finally,
since the CO(4--3) transition observed with APEX has a slightly
smaller beam  than SiO(8--7), we also assumed that all the
SiO(8--7) emission comes from a region as large as the APEX beam at
461\,GHz and corrected the SiO(8--7) emission for dilution.
Under these assumptions, the abundance of SiO varies between $1.2\times 10^{-8}$ and 
$6.8\times 10^{-8}$  with an average value of $2.6\times 10^{-8}$. 
The average value of $X_{\rm{SiO}}$ in our sample increased to $1.8\times 10^{-7}$ when we assume $T_{\rm{rot}}$=5\,K. The derived abundances are plotted against $L_{bol}/M$ 
in Fig.\,\ref{ab}, where G19.61--0.24A is not considered because of the uncertainties on its mass and luminosity. Although this analysis can be performed for only a few sources, they cover the full range of $L_{bol}/M$ values
of our whole sample. We found no
decrease of SiO abundance with $L_{bol}/M$, as reported by \citet{2013A&A...557A..94S} on a
similar sample of sources and based on the SiO(2--1) and (5--4)
transitions. On the other hand, Fig.\,\ref{ab} suggests a rather constant SiO abundance with time given the large
uncertainties in the determination of X$_{\rm{SiO}}$ (which arise because of the local thermodynamic equilibrium assumption and the selected velocity range, for example). 
We note that the trend seen by \citet{2013A&A...557A..94S}   is mostly driven by two sources of their sample
(IRAS\,18151--1208\_1 and IRAS\,19095+0930), while the other sources
show a small variation in $X_{\rm{SiO}}$ compatible with our results.
IRAS\,19095+0930 is also part of our sample  and has the second smallest
abundance of SiO reported in this study.

\begin{table}
\caption[]{SiO abundances in the red-shifted wings.}\label{abu}
\centering
\begin{tabular}{lcrc}
\hline 
\hline
\multicolumn{1}{c}{Source} &
\multicolumn{1}{c}{$\Delta\varv_{\rm{rd}}$} &
\multicolumn{1}{c}{$N_{\rm{SiO}}$} &
\multicolumn{1}{c}{$X_{\rm{SiO}}$}\\

\multicolumn{1}{c}{ } &
\multicolumn{1}{c}{(km\,s$^{-1}$)}&
\multicolumn{1}{c}{($10^{13}$ cm$^{-2}$)}&
($10^{-8}$)\\

\hline
G34.43+0.2M1    &61--83& 3.0$\pm0.2$ & 2.1$\pm0.2$\\ 
IRAS\,18507+0121&69--76& 0.2$\pm0.1$ & 0.7$\pm0.5$\\
IRAS\,19095+0930&52--58& 0.6$\pm0.1$ & 1.2$\pm0.2$\\ 
G19.61--0.24A    &55--85& 4.8$\pm0.3$& 3.8$\pm0.2$\\ 
G20.08--0.14     &52--64& 0.7$\pm0.2$& 4.8$\pm1.1$\\ 
IRAS\,18264--1152&59--107&4.2$\pm0.3$& 3.2$\pm0.2$\\ 

\hline
\end{tabular}
\end{table}

To further test our
results, we also investigated the $L_{\rm{SiO87}}/L_{\rm{bol}}$ ratio as function of
$L_{\rm{bol}}/M$, because \citet{2011A&A...526L...2L} suggested that the variation of SiO luminosity, measured through the (2--1) line, with
$L_{\rm{bol}}/M$ might reflect a change of SiO abundance. Contrary to the result of \citet{2011A&A...526L...2L}, we found no
correlation between $L_{\rm{SiO87}}/L_{\rm{bol}}$ and
$L_{\rm{bol}}/M$. This is probably due to our selection criterion  of broad profiles, however, which
excluded weak SiO(2–1) sources; a similar behaviour was also found in the $L_{\rm{SiO21}}/L_{\rm{bol}}$ 
for the sources we analysed (Fig.\,\ref{ana}).

\subsection{Large velocity gradient calculations}\label{lvg}
We analysed the observed SiO(8--7)/SiO(3--2) (and SiO(8--7)/SiO(5--4) when available) line ratios using the 
 RADEX\footnote{http://home.strw.leidenuniv.nl/$\sim$moldata/radex.html}
non-LTE code \citep{2007A&A...468..627V} with the
rate coefficients for collisions with H$_2$
\citep{2006A&A...459..297D} using a plane parallel geometry. The SiO(5--4) was taken from \citet{2013A&A...557A..94S} and was smoothed over the SiO(8--7) beam to avoid different dilution effects.
We explored H$_2$
densities from $10^2$ to $10^8$ cm$^{-3}$, kinetic temperatures, $T_{\rm kin}$, from
50 to 500 K, and
a specific column density $N$(SiO)/$(dV/dz)$ ranging
from $10^{10}$ to $10^{16}$\,cm$^{-2}$ (km\,s$^{-1}$)$^{-1}$, that is, from the fully
optically thin to the optically thick regime.
No correction for beam dilution was applied, which means that
we implicitly
assumed the same emitting size for the lines. The solutions therefore
are representative of the beam-averaged emission ($\sim$ 18$\arcsec$ for both lines). 

\begin{figure*}
\centering
\subfigure[][]{\includegraphics[width=7.5cm]{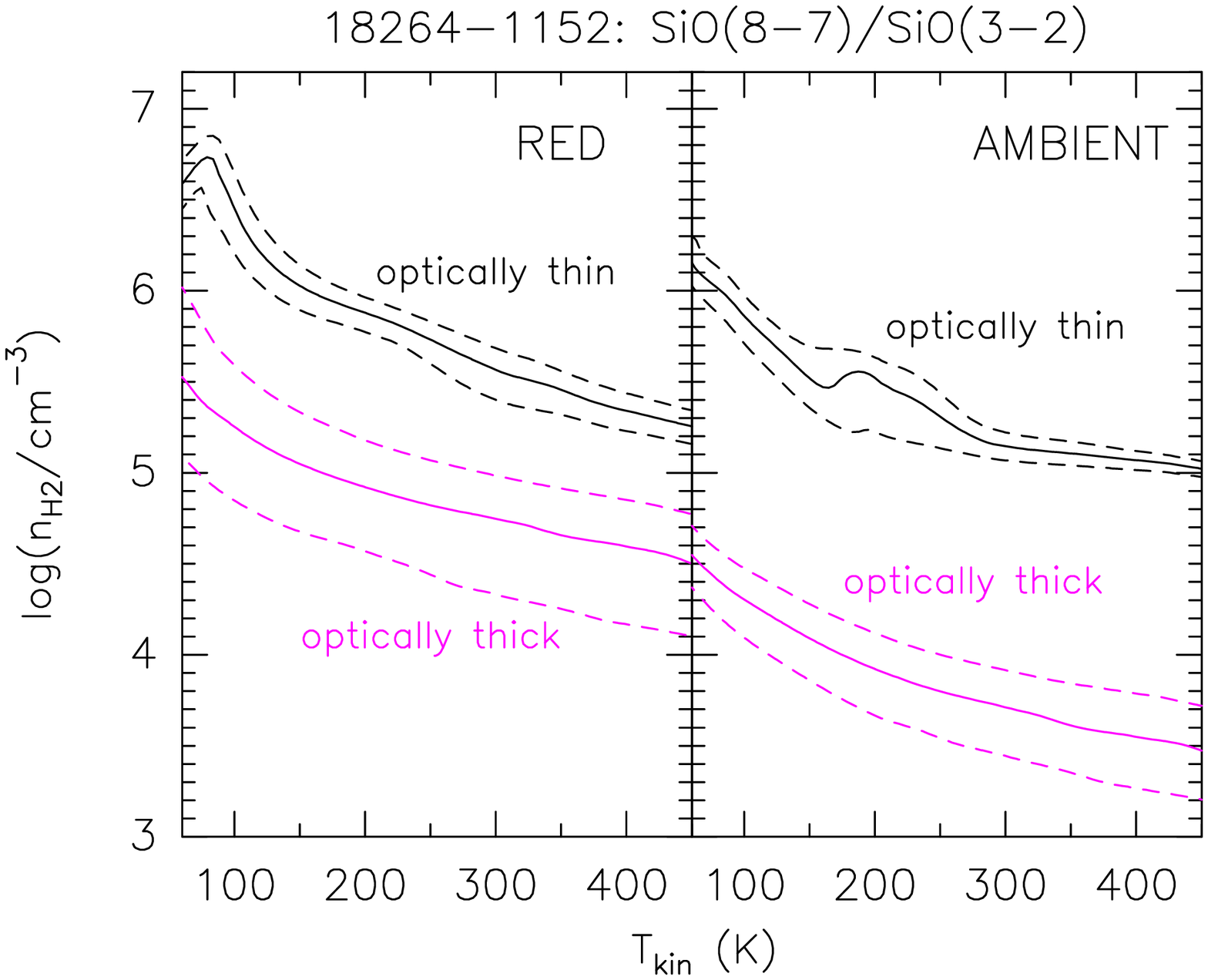}}
\subfigure[][]{\includegraphics[width=10.5cm]{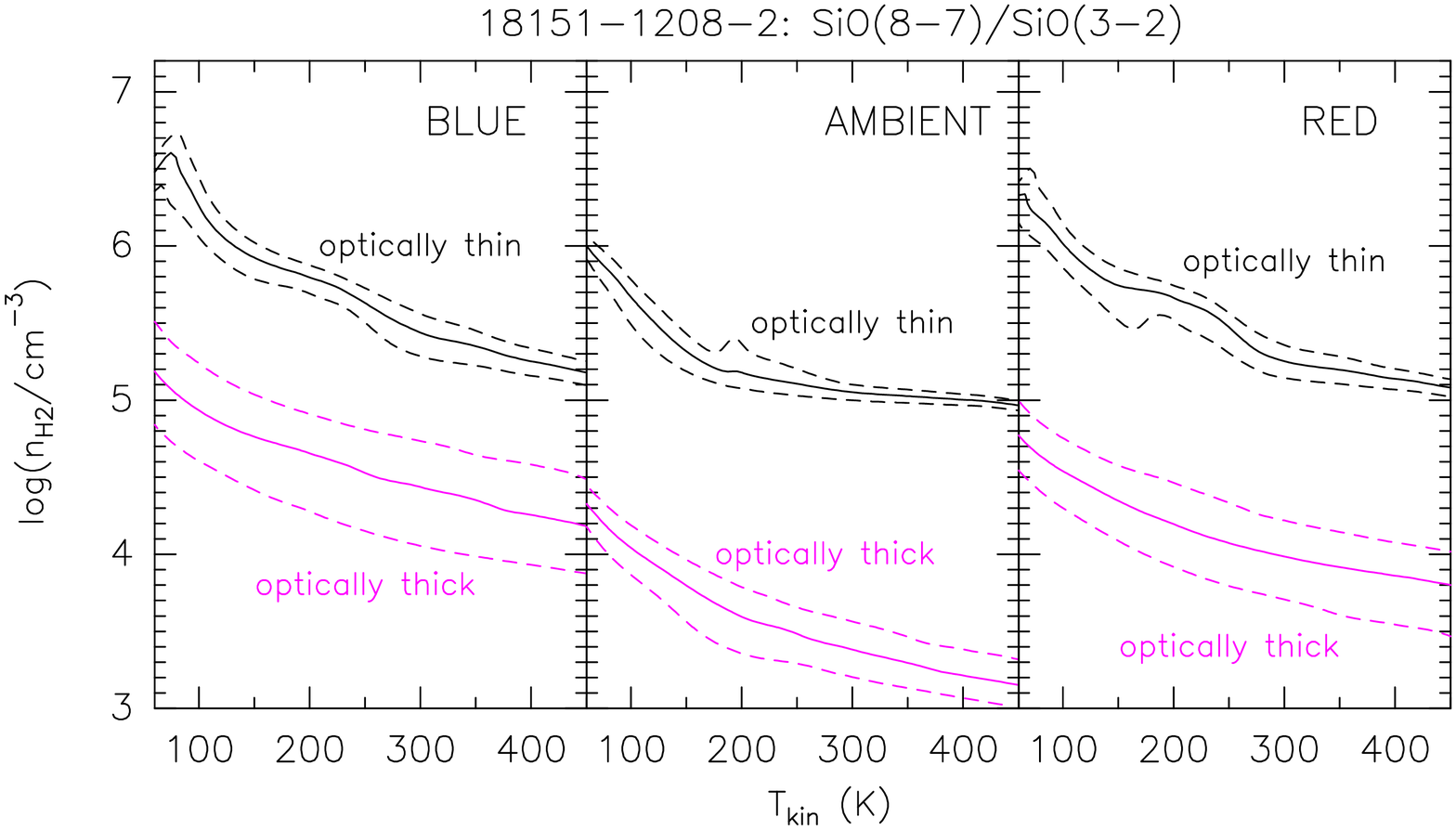}}\caption{Analysis of the SiO
line emission in the IRAS\,18264--1152 (a) and IRAS\,18151--1208$\_$2 (b) outflows
observed with the APEX ($J$ = 8--7, present work) and IRAM 30\,m
($J$ = 3--2, from \citealt{2011A&A...526L...2L}) antennas.
Both lines have been observed with a well-matching beam ($\sim$18$\arcsec$):
the solutions for the observed SiO(8--7)/SiO(3--2) intensity ratio
are shown in the $n_{\rm H_2}$--$T_{\rm kin}$-plane for non-LTE (RADEX)
plane-parallel models for both the optically thin ($N$(SiO) = 10$^{11}$ cm$^{-2}$)
and optically thick ($N$(SiO) = 10$^{16}$ cm$^{-2}$) cases. 
Contours are for the measured ratios
at the typical red- (+54 km s$^{-1}$, for IRAS\,18264--1152; +50 km\,s$^{-1}$, for IRAS\,18151--1208$\_$2) and
blue-shifted outflow 
velocities (+15 km\,s$^{-1}$ for IRAS\,18151--1208$\_$2), as well as at the systemic velocity (IRAS\,18264--1152: +43.9 km s$^{-1}$, IRAS\,18151--1208$\_$2: +30 km\,s$^{-1}$).
Dashed contours are for uncertainties.}
\label{lvg_ratio}
\end{figure*}

\begin{figure}
\centering
\includegraphics[bb = 64 20 530 485,clip,angle=-90,width=7cm]{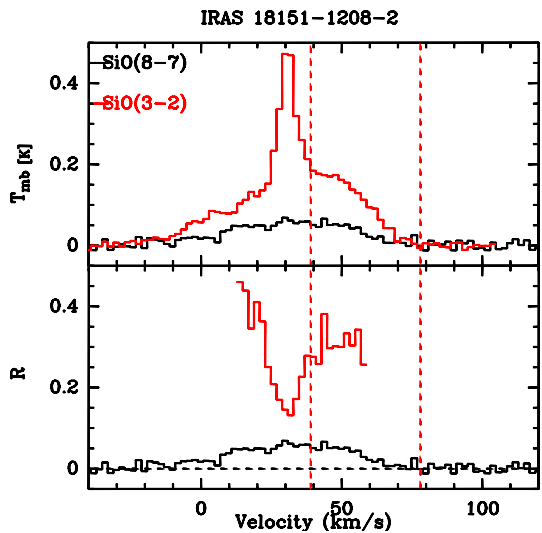}
\caption{{\it Upper panels:}  SiO(3--2) (red) and (8--7) (black) spectra of IRAS\,18151-1208\_2. {\it Lower panels}: SiO(8--7)/SiO(3--2) line ratio as function of velocity (red line). The black line is the SiO(8--7) spectrum. The dashed black line marks the zero $T_{\rm{MB}}$ level of the SiO(8--7) spectra. The dashed red lines label the velocity range used in Sect.\,\ref{velo} for the shock analysis and reported in Table\,\ref{table-ratio}.}\label{87r32}
\end{figure}

Although three transitions are not enough to properly analyse the excitation conditions, 
 useful constraints can be still placed.  For the sources that
were also observed in SiO(5--4), the SiO(8--7)/SiO(3--2) and SiO(8--7)/SiO(5--4) line ratios behave similarly.
As an example, Fig.~\ref{lvg_ratio} shows the solutions for the observed SiO(8--7)/SiO(3--2)
intensity ratio towards two sources of the sample, IRAS\,18264--1152 and IRAS\,18151--1208$\_$2, for typical red-,  blue-shifted, and systemic  
emission in the $n_{\rm H_2}$--$T_{\rm kin}$ plane and for both the optically thin 
($N$(SiO) = 10$^{11}$ cm$^{-2}$) and optically thick ($N$(SiO) = 10$^{16}$ cm$^{-2}$) cases.
As already learnt from previous studies
of low-mass protostellar systems \citep[e.g.,][and references therein]{2007A&A...468L..29C},
the SiO line ratios are not very sensitive to kinetic temperature.
On the other hand, densities always higher than 10$^{3}$ cm$^{-3}$
can be inferred. The average SiO(8--7)/SiO(3--2) line ratios measured at typical red- (0.48$\pm$0.34) 
and blue-shifted 
(0.66$\pm$0.36) outflow velocities ($\vert$$V$--$V_{\rm sys}$$\vert$ $\sim$ 10--15 km s$^{-1}$) 
are higher by a factor $\sim$ 2--3 than those measured 
at the systemic velocity (0.15$\pm$0.10). This trend is better visible in Fig.\,\ref{87r32}, where the
SiO(8--7)/SiO(3--2) line ratio is shown as function of velocity for IRAS\,18151--1208\_2. 
These results show that the excitation conditions of SiO increase with the velocity of the emitting gas, and that
emission at high velocities traces a gas closer to the primary jet, as found 
in
low-mass protostars by \citet{2007A&A...462..163N}. In the red- and blue-shifted 
outflow velocity ranges we find a lower limit to the density of  10$^{4}$ cm$^{-3}$, while at ambient velocities 
the SiO(8--7)/SiO(3--2) line ratio constrains the density to values higher than 10$^{3}$ cm$^{-3}$.
Interestingly, this is consistent with the findings of 
\citet{2007A&A...462..163N} of volume densities between 10$^5$ and 10$^6$ cm$^{-3}$ towards
the nearby prototypical Class 0 low-mass objects L1148 and L1157
 on angular scales of 10$\arcsec$--30$\arcsec$.
Finally, we can obtain lower limits on column densities by 
using the SiO brightness temperatures and assuming the emitting size
$>$ 3$\arcsec$, consistently with the SiO(5--4) maps by \citet{2013A&A...557A..94S},
which shows outflow lobes barely resolved with a beam of 11$\arcsec$. In this way, we obtain  
$N$(SiO) $\geq$ 10$^{13}$--10$^{14}$ cm$^{-2}$ , in agreement with the values derived in Sect\,\ref{xsio}.

\subsection{Shock analysis}\label{velo}

We compared the SiO data with state-of-the-art one-dimensional models of shock propagation in the interstellar medium. We extracted the observed integrated intensity of the shocked gas for both SiO (3--2) and (8--7) lines, and we computed and used the (8--7)/(3--2) integrated intensity ratio. Owing to contamination (Sect.\,\ref{obs87}) of the blue lobe in the SiO (8--7) line for several objects in our sample, we chose to extract these three quantities only from the red-shifted shocked component. Table~\ref{table-ratio} provides the details relevant to this operation for each object, including the velocity range on which the ratio was extracted. The velocity ranges used here differ  from those discussed in Sect.~\ref{xsio} because CO(4--3) is more contaminated by
ambient gas emission than SiO at low-velocities.

\begin{figure}
   \centering
   \includegraphics[width=0.4\textwidth,angle=0]{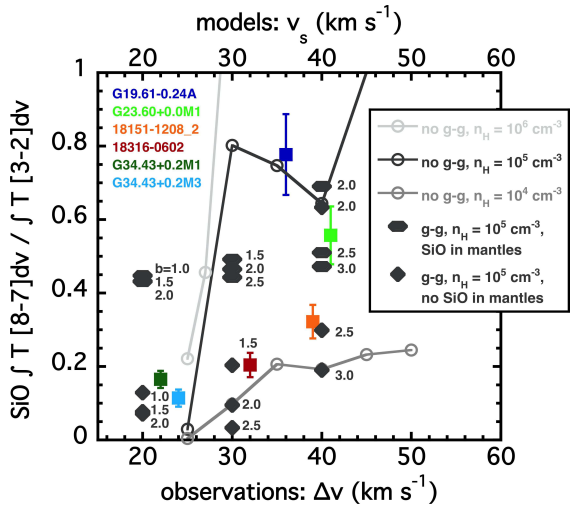}
      \caption{Comparisons of observations and models of SiO (8-7) to (3-2) integrated intensity ratios, extracted over the red-shifted wing of the observed line profiles. For observations, the ratios are displayed against the width of the velocity range over which they are extracted. For models, the ratios are plotted as a function of the shock velocity of the model. In the context of 1D shock modelling, the width of the velocity range used for observations should be directly comparable with the shock velocity of the models for each observed outflow. The observations are shown in coloured squares for various objects in our sample. The empty circles are the models of \citet{Gusdorf081} for pre-shock densities of 10$^4$, 10$^5$, 10$^6$~cm$^{-3}$, the filled diamonds are the models ($n_{\rm H} = 10^5$~cm$^{-3}$) of \citet{Anderl:2013} without SiO initially in the grain mantles, and the filled flat hexagons are the same, including an initial amount of SiO in the grain mantles. 
              }
         \label{figure-ratio}
   \end{figure}

\begin{table*}
\caption{Observed integrated intensity $\int T_{\rm MB} \, d\upsilon$ for the SiO (3-2) and (8-7 lines), corresponding ratio and 
velocity range used for the integration for the shock analysis.}             
\label{table-ratio}      
\centering                          
\begin{tabular}{l   c  c  c c}        
\hline           
object &  $\int T_{\rm MB} \, d\upsilon$ (3-2) & $\int T_{\rm MB} \, d\upsilon$ (8-7) & ratio &$\upsilon_{\rm min}-\upsilon_{\rm max}$ \\
            & (km s$^{-1}$) & (K km s$^{-1}$) & (no unit) & (km s$^{-1}$)\\
\hline
\hline
G19.27+0.1M1   & 1.52$\pm$0.15 & 0.40$\pm$0.04 & 0.26$\pm$0.04& 32 -- 54 \\
G34.43+0.2M1   & 5.40$\pm$0.54 & 0.89$\pm$0.09 & 0.17$\pm$0.02& 61 -- 83 \\   
19095+0930     & 1.91$\pm$0.19 & 0.50$\pm$0.05 & 0.26$\pm$0.04& 46 -- 58 \\   
G19.61-0.24A   & 3.05$\pm$0.30 & 2.37$\pm$0.24 & 0.78$\pm$0.11& 49 -- 85 \\
18264-1152     & 2.93$\pm$0.29 & 1.92$\pm$0.19 & 0.65$\pm$0.09& 47 -- 107\\ 
G20.08-0.14    & 1.84$\pm$0.18 & 0.20$\pm$0.04 & 0.11$\pm$0.03& 52 -- 64 \\ 
18507+0121\_1  & 1.48$\pm$0.15 & 0.26$\pm$0.03 & 0.17$\pm$0.03& 61 -- 76 \\
18316-0602     & 2.80$\pm$0.28 & 0.57$\pm$0.07 & 0.20$\pm$0.03& 53 -- 81 \\
G34.43+0.2M3   & 3.13$\pm$0.31 & 0.36$\pm$0.06 & 0.11$\pm$0.02& 64 -- 88 \\
18151-1208\_2  & 3.87$\pm$0.39 & 1.25$\pm$0.12 & 0.32$\pm$0.05& 39 -- 78 \\
G23.60+0.0M1   & 2.51$\pm$0.25 & 1.40$\pm$0.14 & 0.56$\pm$0.08& 111-- 152\\
\hline
\end{tabular}
\end{table*}
Figure~\ref{figure-ratio} compares the observed ratios  with the values for three classes of 1D stationary shock models: the observations are displayed in colours, while the model points are indicated in shades of grey. We adopted two different x-axes: for the models, the x-axis is the shock velocity,  for the observations
it is the width of the velocity range used for the integration. When trying to fit observations with 1D models, these two quantities should be of the same order. We subsequently chose to restrict the analysis to observations for which the velocity  range used for the observations could be compared with the shock velocity of our models, namely between 20 and 50~km~s$^{-1}$. The three model classes shown in Fig.\,\ref{figure-ratio} are:
\begin{itemize}
\item class 1 models   from \citet{Gusdorf081}: pre-shock densities of 10$^4$, 10$^5$, 10$^6$~cm$^{-3}$, magnetic field parameter\footnote{The magnetic field parameter $b$ relies on the assumption that the magnetic field strength, $B$, in dense clouds scales with the total proton density, $n{}_{\rm H}$, as $B(\mu {\rm G})=b\sqrt{n_{\rm H}({\rm cm}^{-3})}$ \citep{Crutcher:1999}.} $b$ = 1, and Si-bearing material only distributed in the grain cores in the pre-shock phase;
\item class 2 models  from \citet{Anderl:2013}: they include grain-grain interactions. Three different shock velocities (20, 30 and 40 km s$^{-1}$) are covered, with three different values of the magnetic field for each velocity ($b=1.0, 1.5, 2.0$ for 20 km\,s${}^{-1}$, $b=1.5, 2.0, 2.5$ for 30 km\,s${}^{-1}$, and $b=2.0, 2.5, 3.0$ for 40 km\,s${}^{-1}$) at a pre-shock density of $10^5\, $cm$^{-3}$. Compared with models that do not incorporate grain-grain processing, higher values of the magnetic field parameter $b$ are required for continuous magneto-hydrodynamic shocks to propagate at higher velocities because of the different coupling of small charged grains to the magnetic field;
\item class 3 models  are of unprecedented kind: they are similar to class 2 models, but additionally, 10\% of the elemental Si is assumed to be in the form of SiO in the grain mantles in the pre-shock phase. One of the important conclusions in \citet{Anderl:2013} is that these models might be the only way to fit SiO observations of outflows from low-mass YSOs. For this class, the range of covered input parameters is the same as for class 2. 
\end{itemize}

In a first step, we compared the observations with the class 1 models, in which SiO cannot be efficiently produced with shock velocities lower than 25~km~s$^{-1}$ because this is the threshold for core erosion to efficiently release Si in the gas phase. These models unambiguously indicate that only a pre-shock density greater than 10$^4$~cm$^{-3}$ could yield solutions compatible with the observed ratios.
At such high densities, however, the effects of grain-grain interactions (mostly shattering and vaporisation, \citealt{Guillet11}) must be taken into account \citep{Anderl:2013}. These effects are not considered in the class 1 models. 

In class 2 and 3 models, SiO can be released into the gas phase by gas-grain interactions, leading to the sputtering of the grain cores (as in \citealt{Gusdorf081}), but also through grain-grain collisions leading to the vaporisation and shattering of the grains. The latter effect results in a production of small grain fragments in large numbers, which increases the total dust grain surface area and thereby changes the coupling between the neutral and the charged fluids within the shocked layer. The corresponding shocks become much hotter and subsequently thinner, because the kinetic energy of the shock is transformed into heat within a shorter time. In addition to this release of SiO by vaporisation, this strong effect of shattering on the structure of magneto-hydrodynamic shocks is the second reason why it is necessary to include grain-grain processing in molecular shocks at densities higher than $\sim10^5\, $cm$^{-3}$ . In class 3 models, part of the SiO can additionally originate from the sputtering of the grain mantles. 
Two behaviours can be identified:
\begin{itemize}
\item all class 2 models can provide an overall decent fit to most observations in terms of integrated intensity ratios (except for G19.61-0.24A, which might be associated to a higher pre-shock density). However, for a shock velocity equal to 20~km~s$^{-1}$, or for the shock at 30~km~s$^{-1}$ and b\,=\,2.5, the SiO absolute integrated intensities are simply too low to match the observed values assuming an emitting size $>$3\arcsec (see the end of Sect.\,\ref{lvg}). This is because at low shock velocities, the grain-grain interactions are not efficient, and because the combination of a moderate (30~km~s$^{-1}$) shock velocity and high magnetic field strength results in a critically narrow SiO emission layer;
\item the class 3 models do provide convincing fits in terms of absolute integrated intensities, but they seem to generally overestimate the integrated intensity ratios as displayed in Fig.\,\ref{figure-ratio} (except for G19.61-0.24A and G23.60+0.0M1). This is because we chose to put 10\% of SiO in the grain mantles, similar to \citet{Gusdorf082}. Figure\,\ref{figure-ratio} and the absolute integrated intensity criterion suggest that models with fewer SiO in the grain mantles would provide the best fit. 
\end{itemize}

\section{Discussion}
Under the assumption that the SiO emission arises from gas closely associated with the primary jet, as found in molecular jets from low-mass YSOs, our observations confirm the presence of jets up to $L_{\rm {bol}}/M$ values at least of  50 $L_\odot/M_\odot$.
The results presented in Sect.\,\ref{time} 
suggest that excitation conditions of the gas responsible for
the SiO emission increase with time.  The trend of SiO luminosity and luminosity ratios as a function of 
$L_{{\rm bol}}/M$ are weak and similar observations on a larger sample of sources need to be performed to confirm our findings. However, these trends might explain the non-detection of SiO(8--7) in the youngest sources of our sample.
Large velocity gradient calculations showed that the SiO(8--7) to SiO(3--2) line ratio depends more strongly on
density than on temperature.  
This suggests that the increase of SiO(8--7) to SiO(3--2)
seen in our sources reflects an increase of density of the gas in the jet. 
 We speculate that this increase in jet density might be 
linked to an increased
 mass-accretion rate at very early stages of star formation \citep[as noted for HH212 by][]{2007A&A...468L..29C}. In this
 case, the increase of density seen in our sample of sources might be  interpreted as an increase of the accretion rate during the
 evolution of a massive YSO or, alternatively, as an increase of the
 accretion rate with the mass of the central object. Indeed, an increase of  
$L_{bol}/M$ is always linked to an increase of mass of the central object when accretion is still significant, when
$L_{bol}/M$ is either interpreted as a time estimator, or when
it mostly reflects the mass of the most massive object in the cluster.
 Recently, \citet{2013ApJ...772...61K}  modelled the simultaneous evolution of a
 massive protostar and its host molecular core. Their calculations
 showed that the accretion rate increases with time and with the mass of
 the protostellar mass in the very early stages of evolution until the
 beginning of the Kelvin-Helmholtz contraction.   
A note of caution here is
 that effects of radiative pumping processes are not considered in our
 analysis, while \citet{2013A&A...550A...8G}  showed that SiO
 $J \ge 3$ lines are sensitive to optical/UV radiation fields. This
 effect becomes more relevant with time, especially for high $J$
 transitions, as a consequence of the stronger radiation field from the central object. 
However, this mechanism probably mostly affects gas close
 to the YSO due to extinction, unless the SiO emission
 comes from the walls of the outflow cavities that allow the
 radiation to reach longer distances from the protostar. 

Alternatively,  when no pre-shock SiO is placed in the grain mantles, and all other shock model parameters are equal, \citet{Gusdorf082} predicted that the integrated
intensity of  SiO transition tends to increase
with the age of the shock in outflows from low-mass Class\,0
objects, owing to the greater velocity
extent of the emitting region. To verify this, we looked for correlations
between the luminosity of the SiO(8--7), (3--2) and (2--1) transitions and their
line width (Fig.\,\ref{dvs}). The luminosity of the SiO(8--7) increases for
line widths between 30 and 60\,km\,s$^{-1}$, although two objects (IRAS\,1824--152 and G23.60+0.0M1 ) do not follow the general trend.   The SiO(3--2) luminosity (and also that of the SiO(2--1) line) neither shows any strong correlation with its corresponding line width, 
 nor is there an obvious correlation between the line width of the SiO(3--2) and (8--7)
transitions and $L_{\rm{bol}}/M$ because the
line widths are relatively constant in our sample (see
Table\,\ref{sio87}), which probably is a consequence of our selection criteria. On the other hand, the SiO(2--1) line width seems to
decrease with increasing values of $L_{\rm{bol}}/M$.
We note, however, that the observed line width strongly depends on  the inclination angle of the outflow with
respect to the sky, which is  unknown for the sources of our sample.

The second result of our study is that we do not find any significant 
variation of SiO abundance in our sources. This is estimated through comparisons with CO column densities
obtained by mean of the CO(4--3) integrated intensity when available, but also through $L_{\rm{SiO}87}/L_{bol}$ for all the sources in the sample.This result contradicts the findings
of other authors \citep{2006A&A...460..721M,2011A&A...526L...2L,2013A&A...557A..94S}, who reported  a decrease of SiO abundance with
  time.  However, there are two differences between our and the previous studies. First, previous results from the literature were
  obtained through observations of the SiO(2--1) and (3--2) lines,
  which are not sensitive to changes of excitations. Second, 
our study does not sample the most evolved sources where the SiO abundance is expected to drop more dramatically 
because of a decay of jet activity. To investigate the variation of SiO abundance as function of time and of excitation in more
detail, we also need observations of high $J$ SiO transitions towards more evolved objects than those presented here. Moreover,  a more accurate analysis of the SiO emission requires high angular
resolution observations to resolve the emitting gas and verify that
the SiO emission is dominated by one object and does not arise from a
population of lower mass YSOs.

 Finally, our analysis (Fig.\,\ref{figure-ratio}) shows that shocks
 with a rather high pre-shock density ($n_{\rm H} =$10$^5$~cm$^{-3}$)
 provide the best fit to the observations. The analysis shows that
 grain-grain interactions must be taken into account, and that part of
 the SiO emission arises from the sputtering of a fraction of
 silicon-bearing material (corresponding to less than 10\% of the
 total silicon abundance) from the grain mantles. In this sense, we
 corroborate the findings of \citet{Anderl:2013}. Silicon-bearing
 material in the mantles of grains around massive YSOs might be due to previous shock processing of the
 regions, either owing to previous episodes of ejection in the
 targeted regions (as for the B1 knot of the L1157 outflow, invoked by
 \citealt{Gusdorf082}), or to converging flows that might lead to the
 formation of massive stars in filamentary structures (\lq ridges',
 like in the W43 region, e.g.,
 \citealt{2013ApJ...775...88N}). Unfortunately, this shock modelling
 cannot be used for evolutionary trend studies because the shocks are
 stationary. However, dense shocks can account for the observations of
 SiO in massive star-forming regions, as was also reported by
 \citet{2013A&A...554A..35L}.

\section{Conclusions} 
We presented SiO(8--7) and CO(4--3) APEX observations of a sample of massive clumps in different evolutionary phases of star formation. The data were complemented with SiO(3--2) (and SiO(5--4) when available) observations from the literature with a similar angular resolution.
The main results can be summarised as follows:
\begin{enumerate}

\item We detected SiO(8--7) emission in all sources of our sample except in three objects that are probably in very early phases of star formation.
 The line-profiles are very broad, with an average $FWZP$ of 45\,km\,s$^{-1}$.
\item The SiO(8--7) luminosity and the SiO(8--7) to SiO(3--2) line ratio increase weakly with the evolutionary phase of the source described by $L_{bol}/M$. This might be explained with an increase of density in the gas traced by SiO.
\item We analysed the SiO(8--7) to SiO(3--2) line ratio with a non-LTE radiative transfer code. At high outflow velocities we found a lower limit to the density of  10$^{4}$ cm$^{-3}$ , in agreement with measurements  towards
 nearby prototypical Class 0 low-mass objects. We also found that the excitation conditions of SiO increase with the velocity of the emitting gas. This suggests that emission at high velocities traces a gas close to the primary jet, as found  in low-mass protostars.
\item We estimated the abundance of SiO at high velocities through comparison with CO(4--3) emission in the same velocity range as SiO. Although the estimates are affected by large uncertainties, the SiO abundance is realtively constant,
with an average value  of 2.6\,10$^{-8}$ and no clear change of $X_{\rm{SiO}}$ versus $L_{bol}/M$.

\item Shock modelling reveals high pre-shock densities ($n \geq 10^5\, $cm$^{-3}$). At these densities the effects of grain-grain collisions cannot be ignored anymore. For the first time, the models of \citet{Anderl:2013} of magnetohydrodynamical shocks including the effects of shattering and vaporisation were used to interpret observations. The observations are globally compatible with these models at a pre-shock density of $10^5\, $cm$^{-3}$, in which sputtering of silicon-bearing material (corresponding to less than 10\% of the total silicon abundance) from the grain mantles occurs.

\end{enumerate}

\Online

\begin{appendix}
\section{Figures}\label{figures}

\begin{figure*}
\begin{tabular}{c}
\includegraphics[bb = 71 33  529 386,clip,angle=-90,width=6cm]{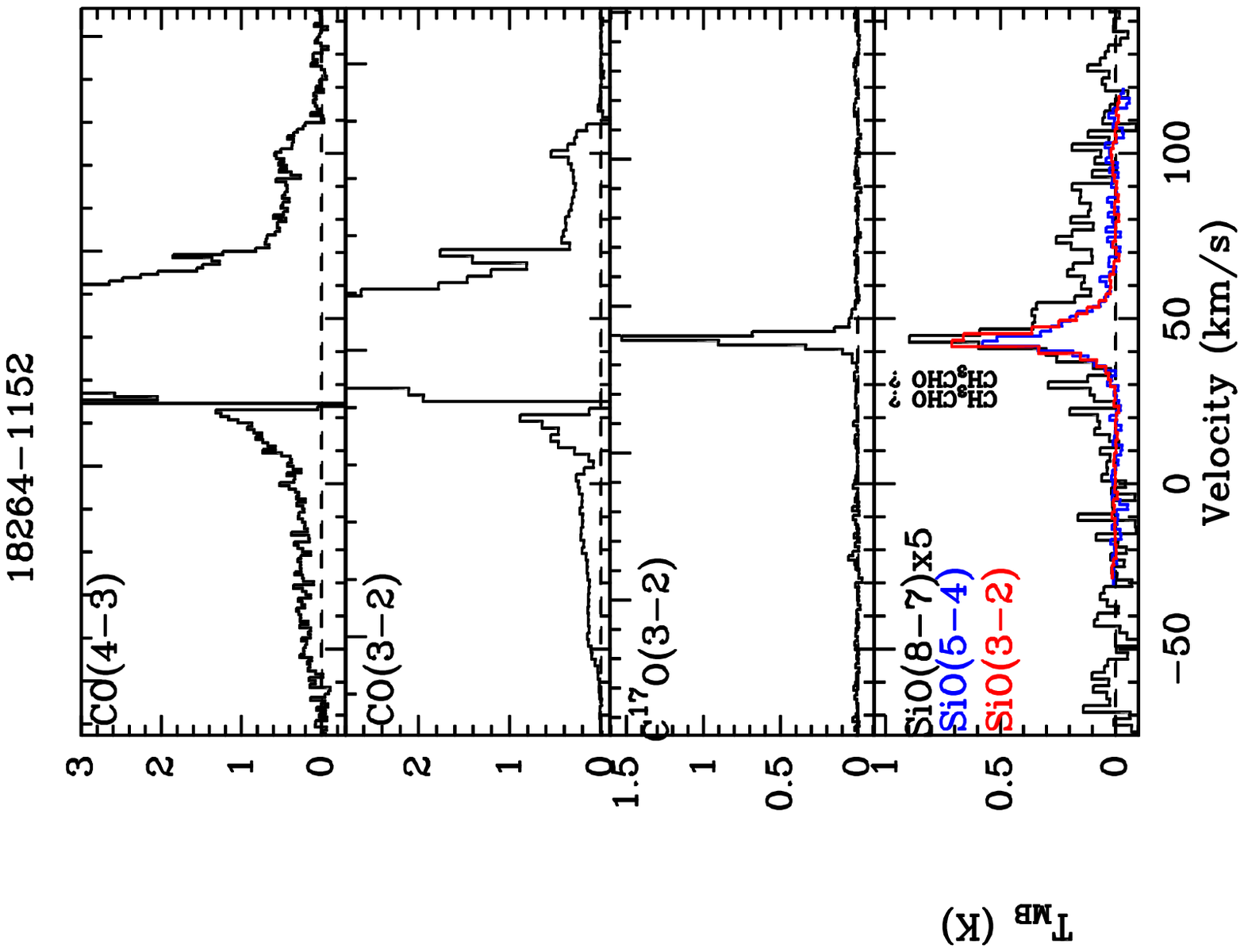}
\includegraphics[bb = 71 33  529 386,clip,angle=-90,width=6cm]{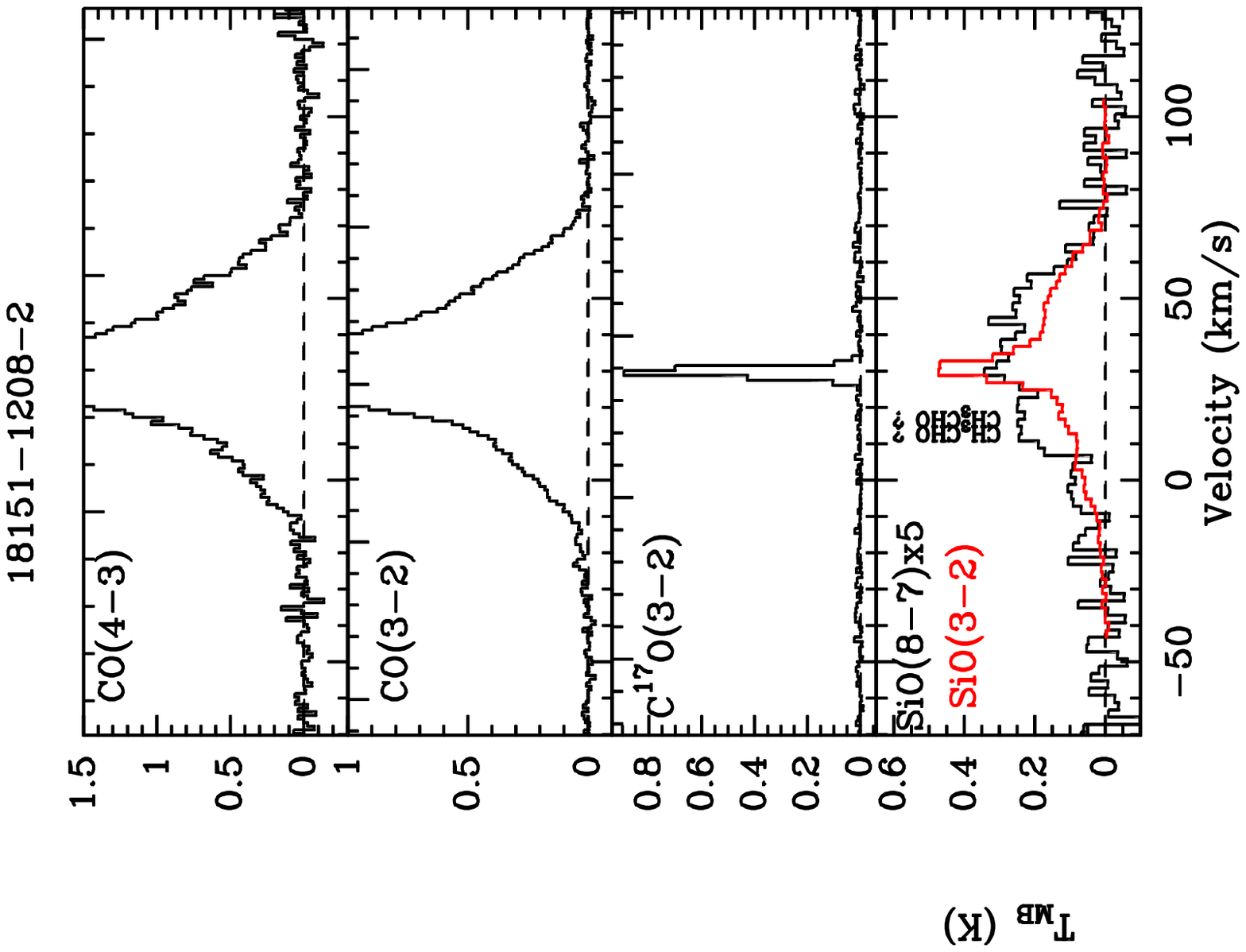}
\includegraphics[bb = 71 33  529 386,clip,angle=-90,width=6cm]{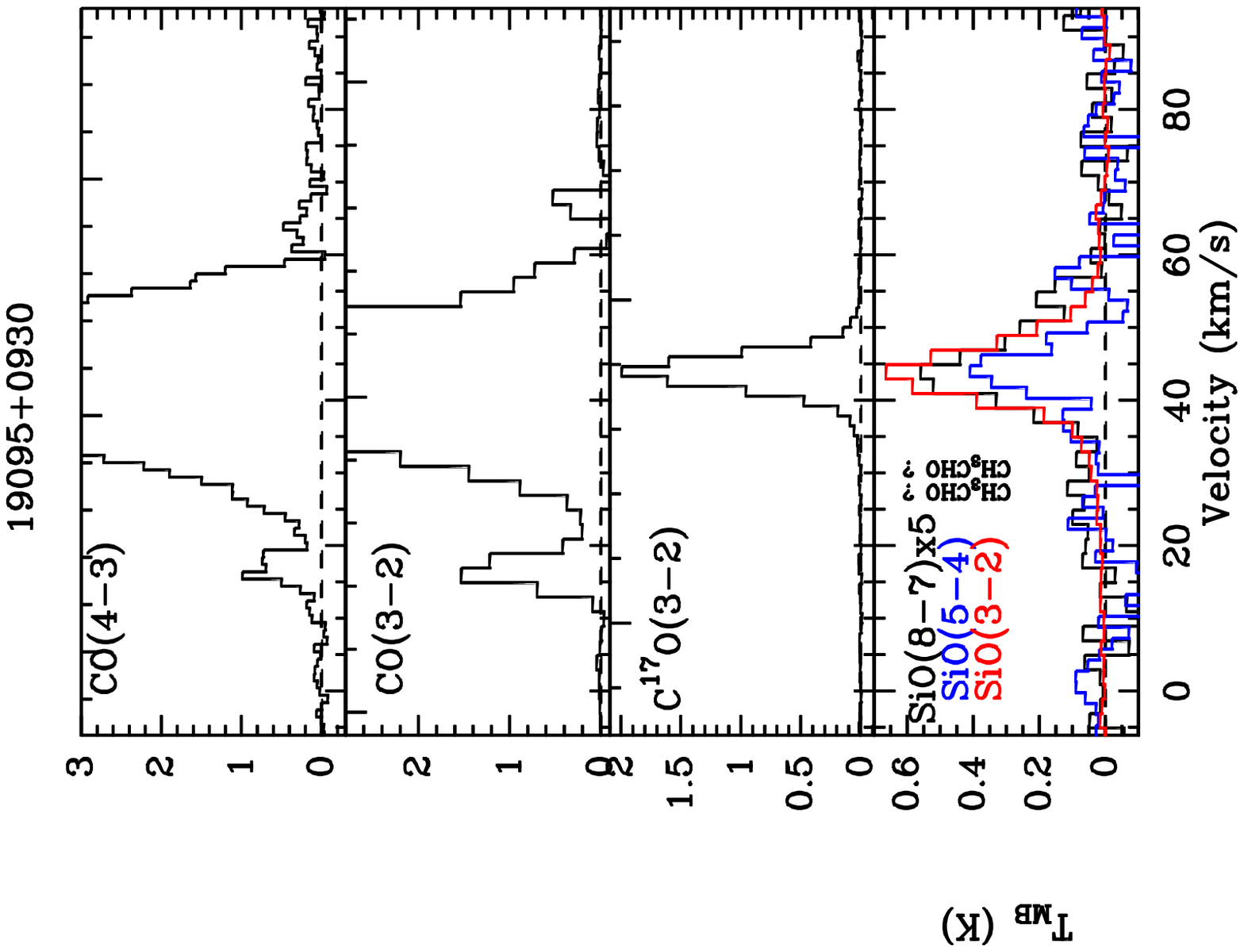}\\
\includegraphics[bb = 71 33  529 386,clip,angle=-90,width=6cm]{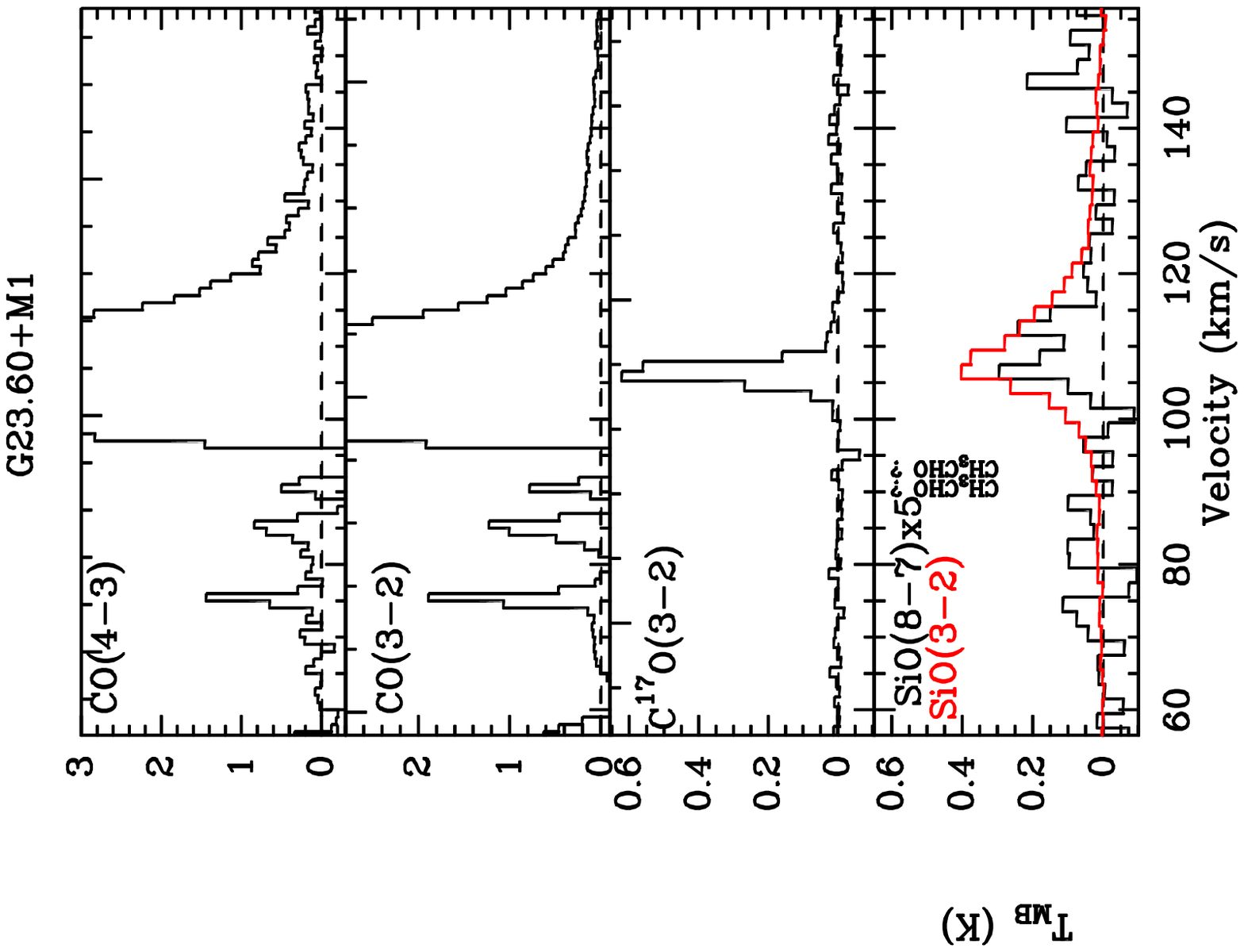}
\includegraphics[bb = 71 33  529 386,clip,angle=-90,width=6cm]{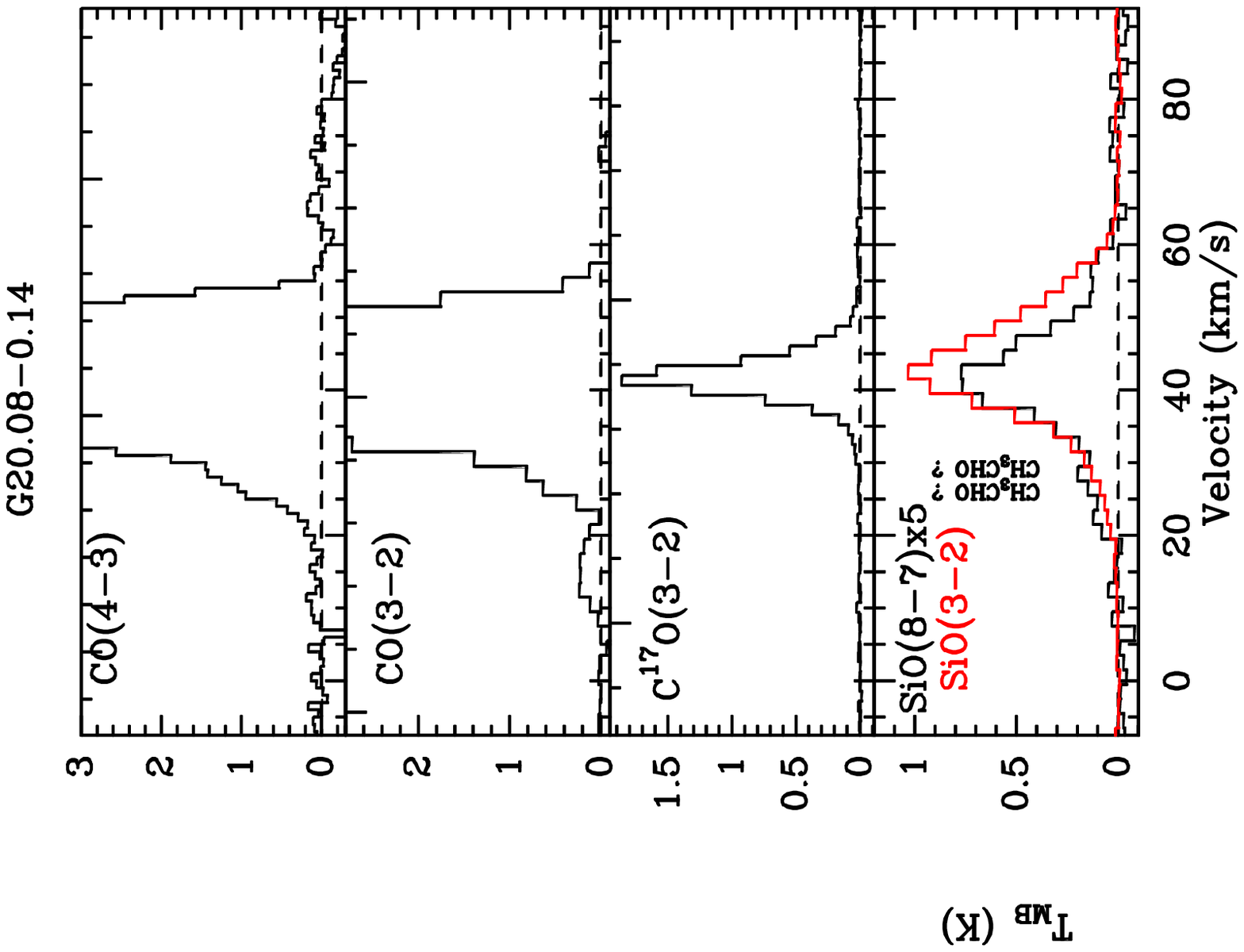}
\includegraphics[bb = 71 33  529 386,clip,angle=-90,width=6cm]{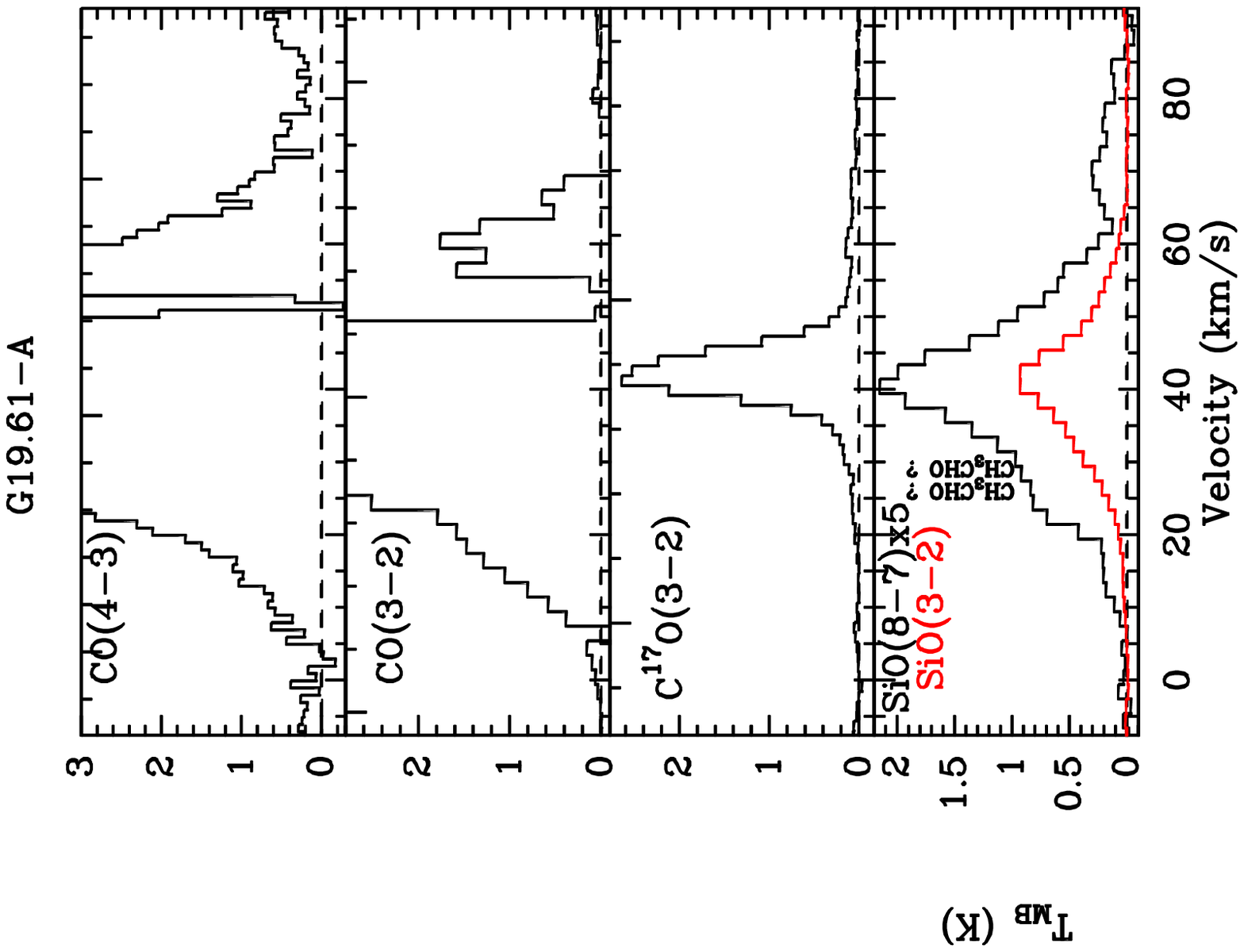}\\
\includegraphics[bb = 71 33  529 386,clip,angle=-90,width=6cm]{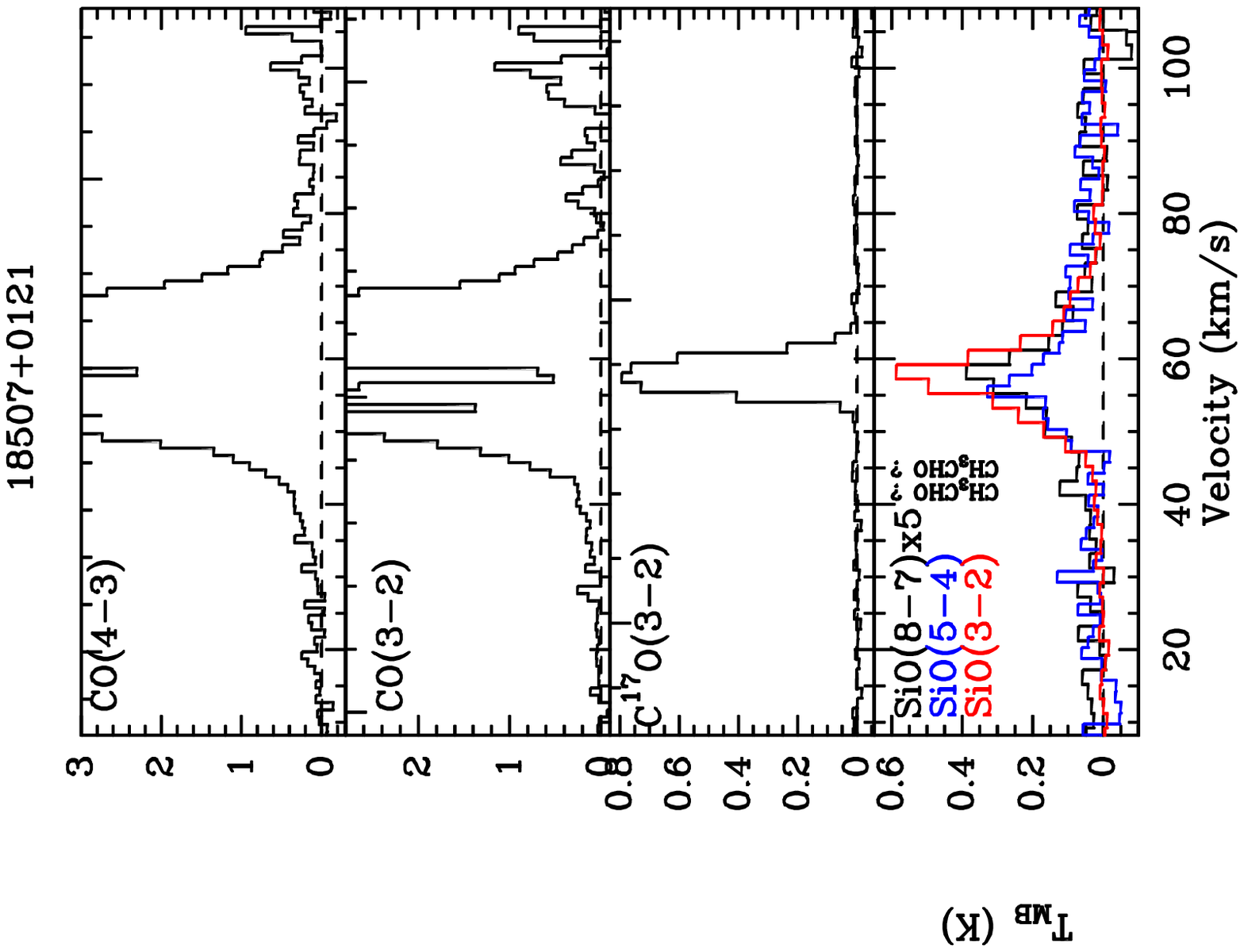}
\includegraphics[bb = 71 33  529 386,clip,angle=-90,width=6cm]{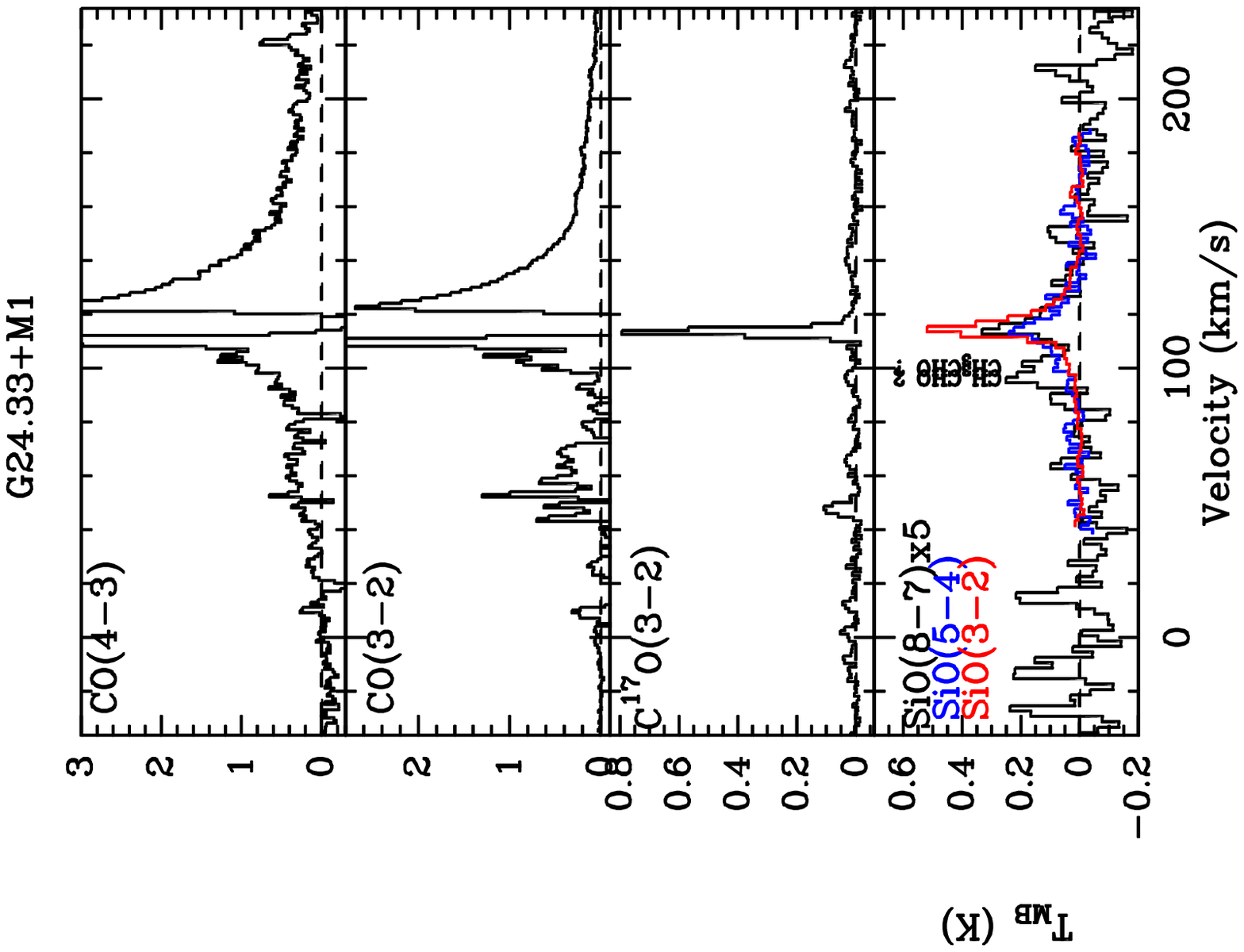}
\end{tabular}
\caption{Spectra of CO(4-3) and (3-2), C$^{17}$O(3-2), SiO(8-7) (multiplied by a factor of five for clarity), SiO(5--4) (blue, where available) and SiO(3--2) (red). The CH$_3$CHO features at 347.35\,GHz (Sect.\,\ref{sio87}) are labelled.\label{spectra1}}
\end{figure*}

\begin{figure*}
\begin{tabular}{c}
\includegraphics[bb = 71 33  529 386,clip,angle=-90,width=8cm]{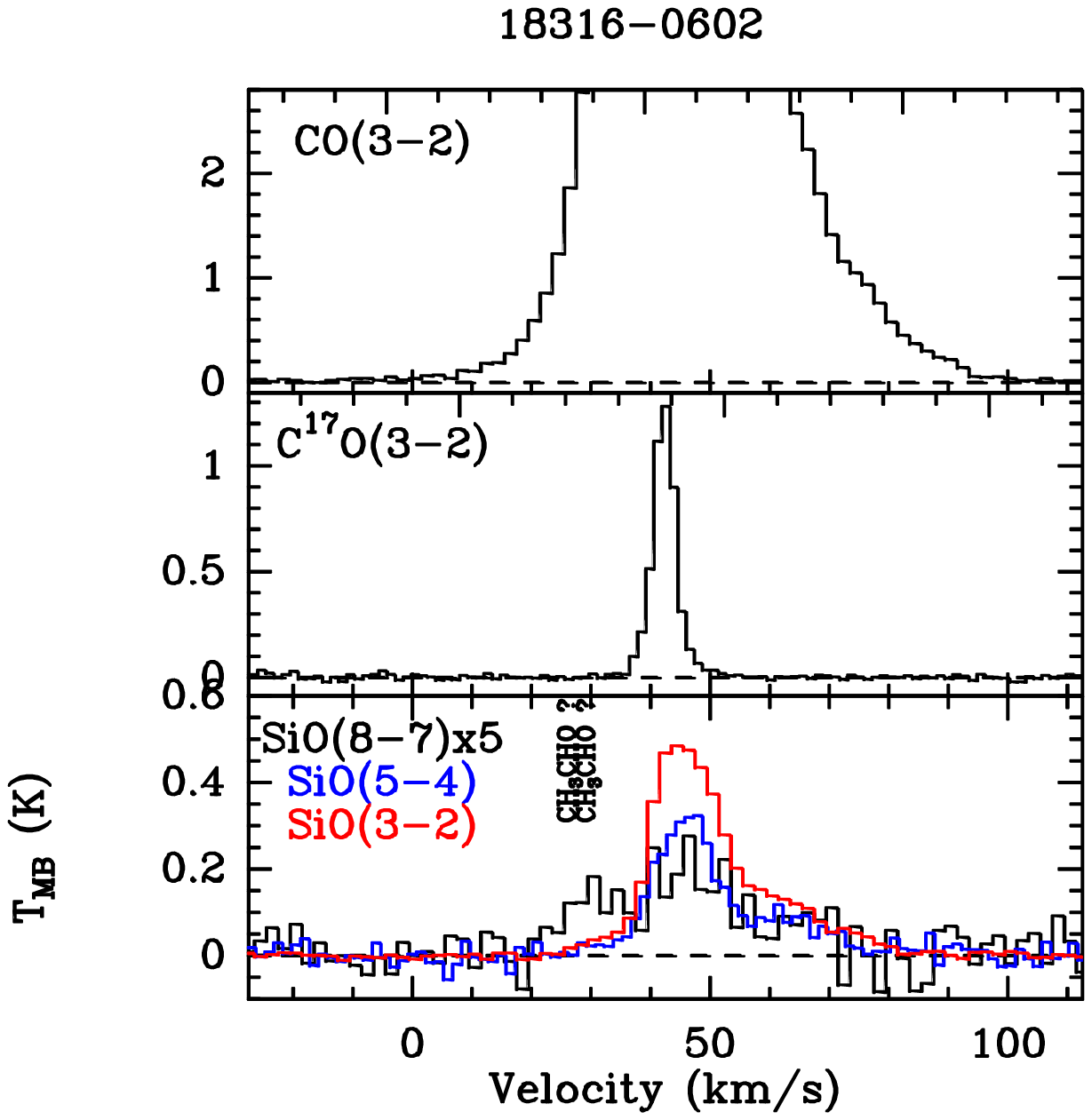}
\includegraphics[bb = 71 33  529 386,clip,angle=-90,width=8cm]{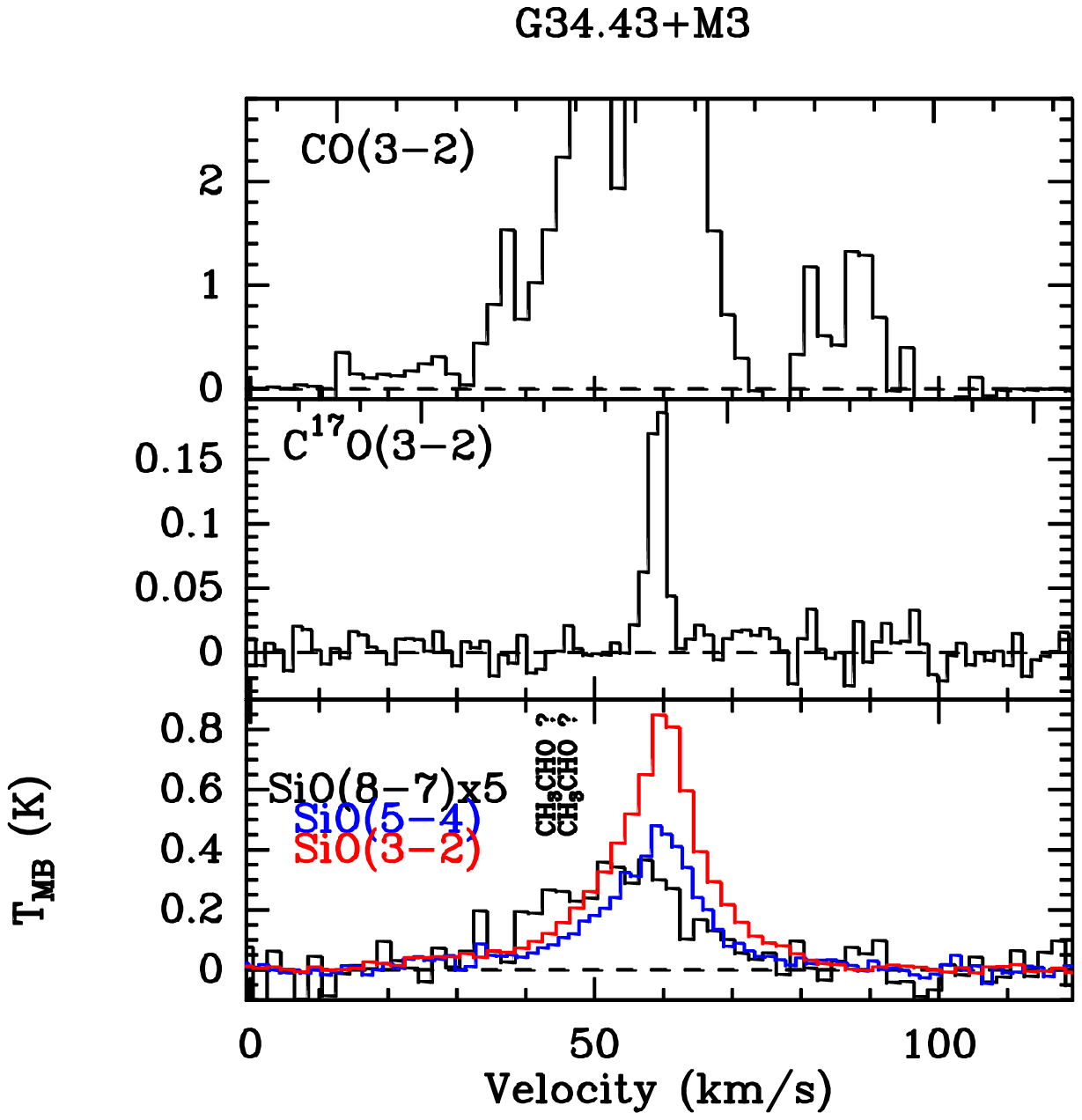}
\end{tabular}
\caption{Spectra of CO(3-2), C$^{17}$O(3-2), SiO(8-7) (multiplied by a factor of five for clarity), SiO(5--4) (blue, where available) and SiO(3--2) (red). The CH$_3$CHO features at 347.35\,GHz (Sect.\,\ref{sio87}) are labelled.\label{spectra2}}
\end{figure*}

\begin{figure*}
\begin{tabular}{c}
\includegraphics[bb = 71 33  529 393,clip,angle=-90,width=6cm]{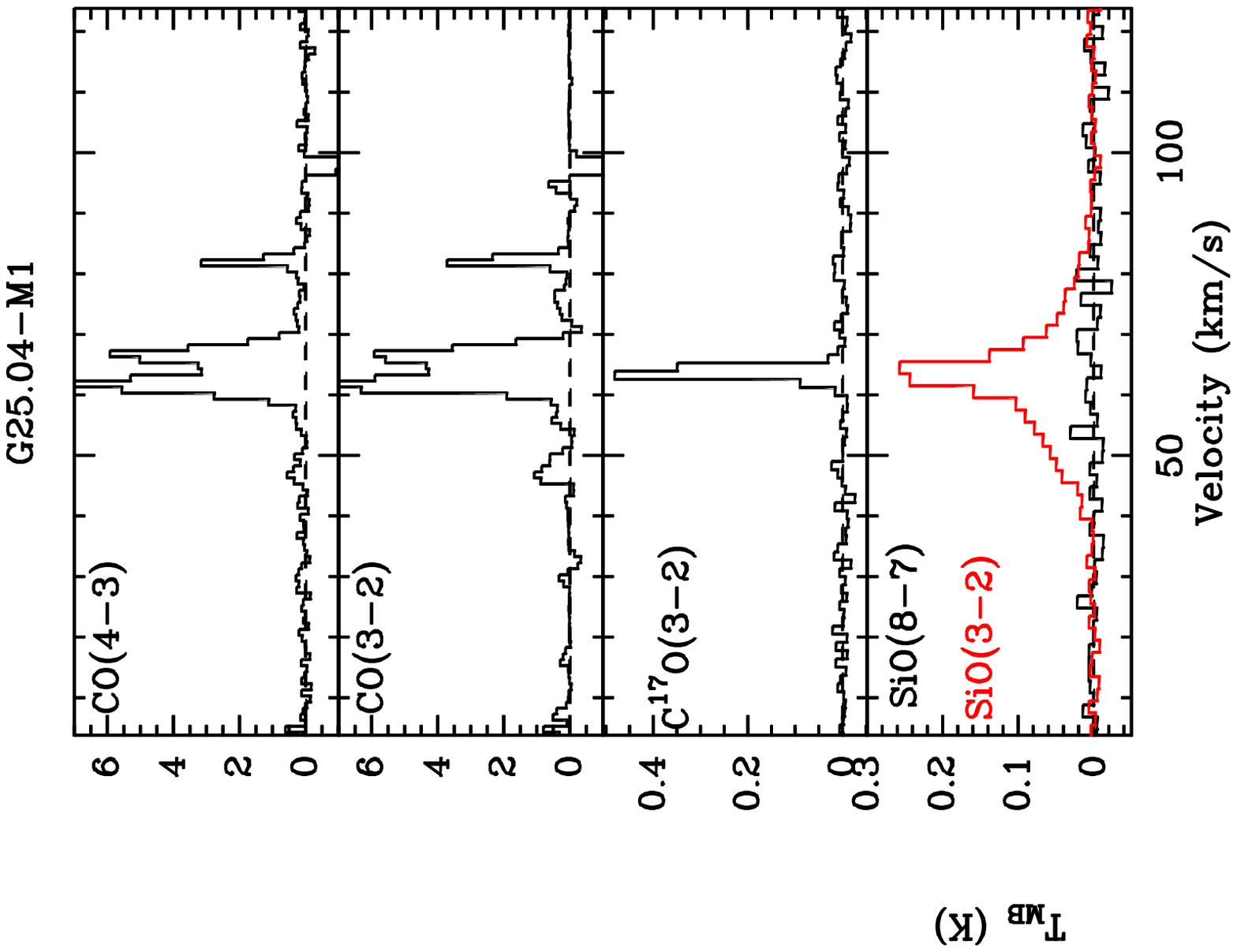}
\includegraphics[bb = 71 33  529 393,clip,angle=-90,width=6cm]{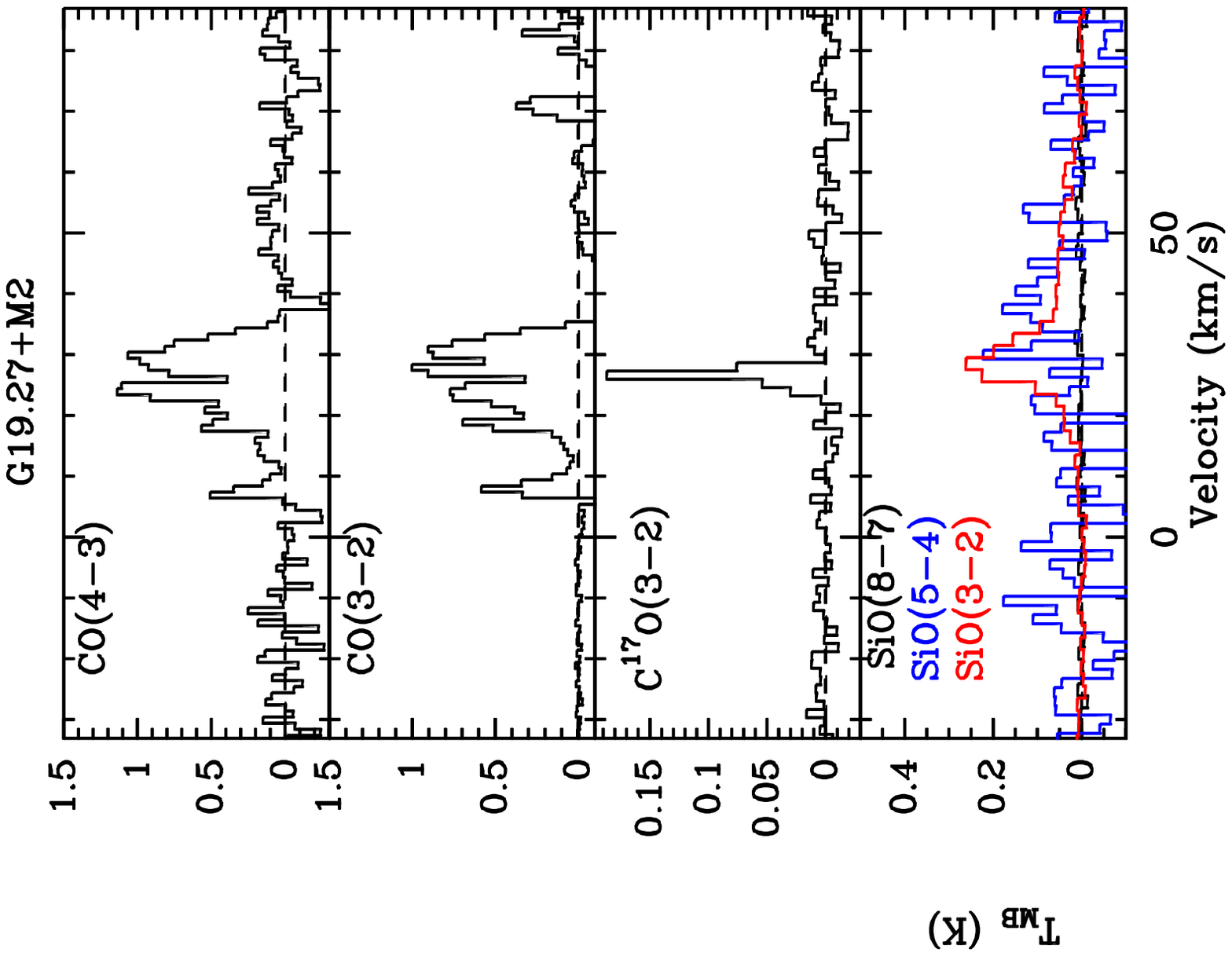}
\includegraphics[bb = 71 33  529 393,clip,angle=-90,width=6cm]{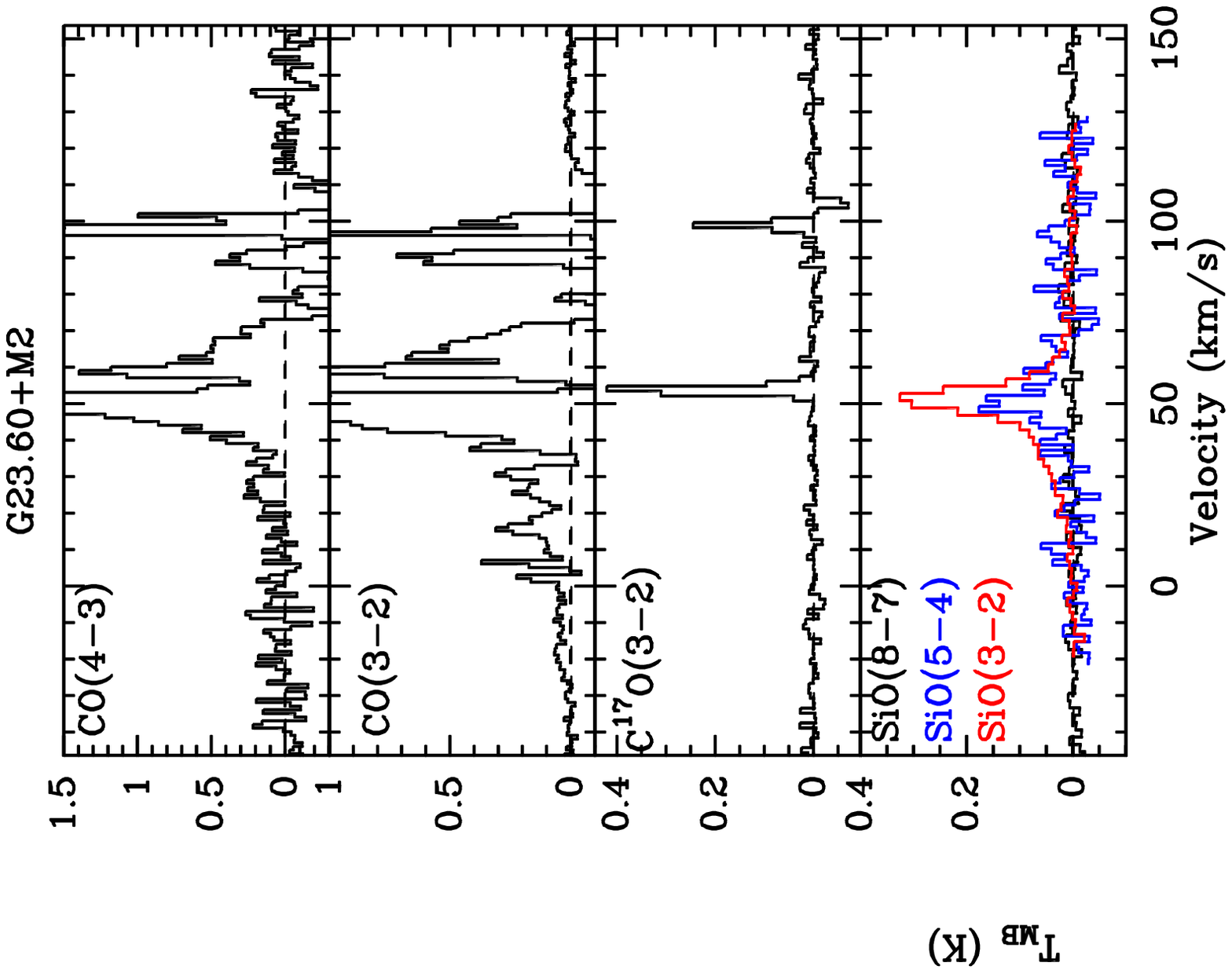}
\end{tabular}
\caption{Spectra of CO(4--3) and (3-2), C$^{17}$O(3-2), SiO(8--7) (black), SiO(5--4) (blue, where available) and SiO(3--2) (red) for the three sources not detected in SiO(8--7) (see Table\,\ref{sio87}).}\label{nondet}
\end{figure*}

\begin{figure*}
\centering
\subfigure[][]{\includegraphics[bb = 88 185 535 698,clip,angle=-90,width=6cm]{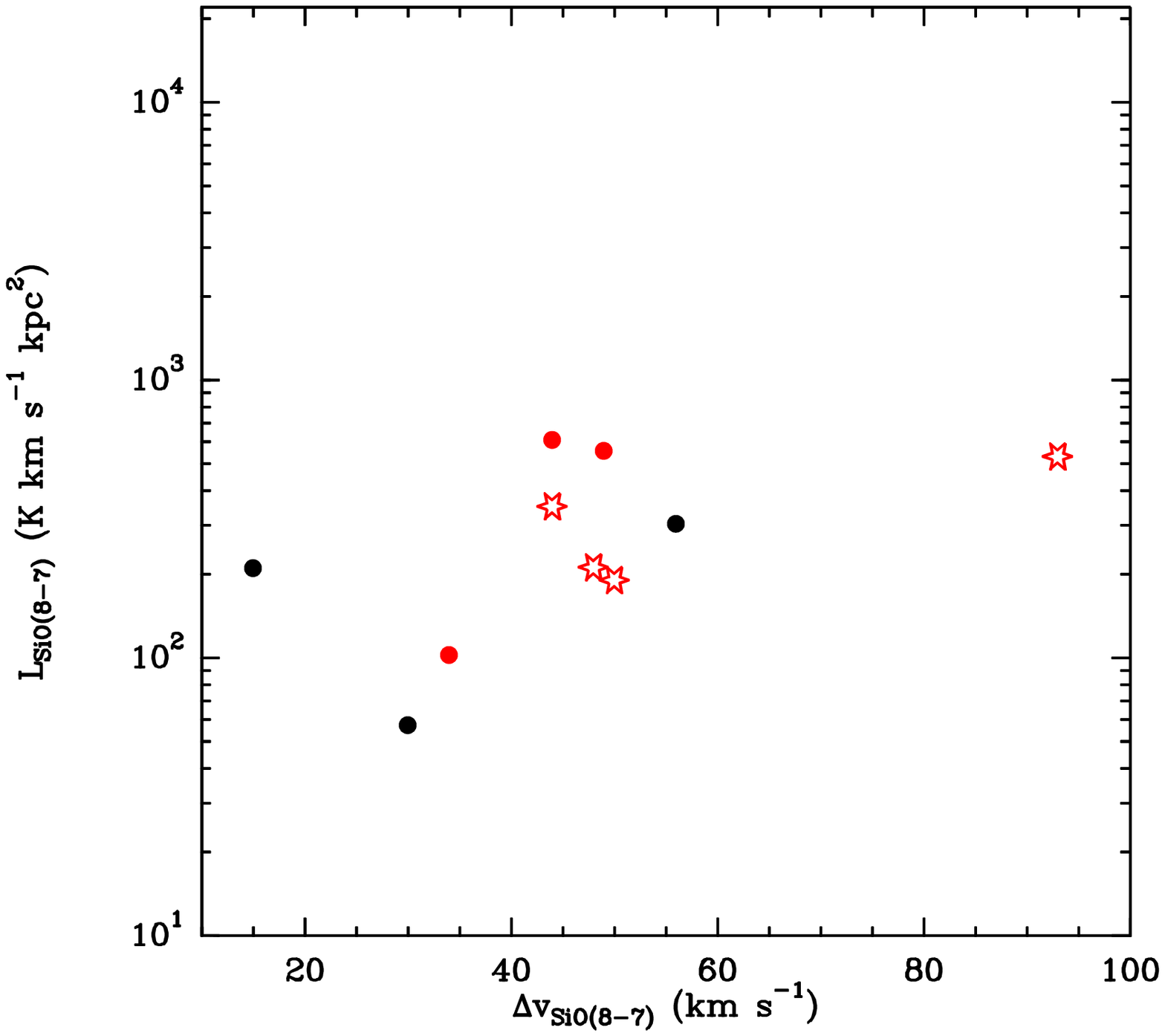}}
\subfigure[][]{\includegraphics[bb = 88 185 535 698,clip,angle=-90,width=6cm]{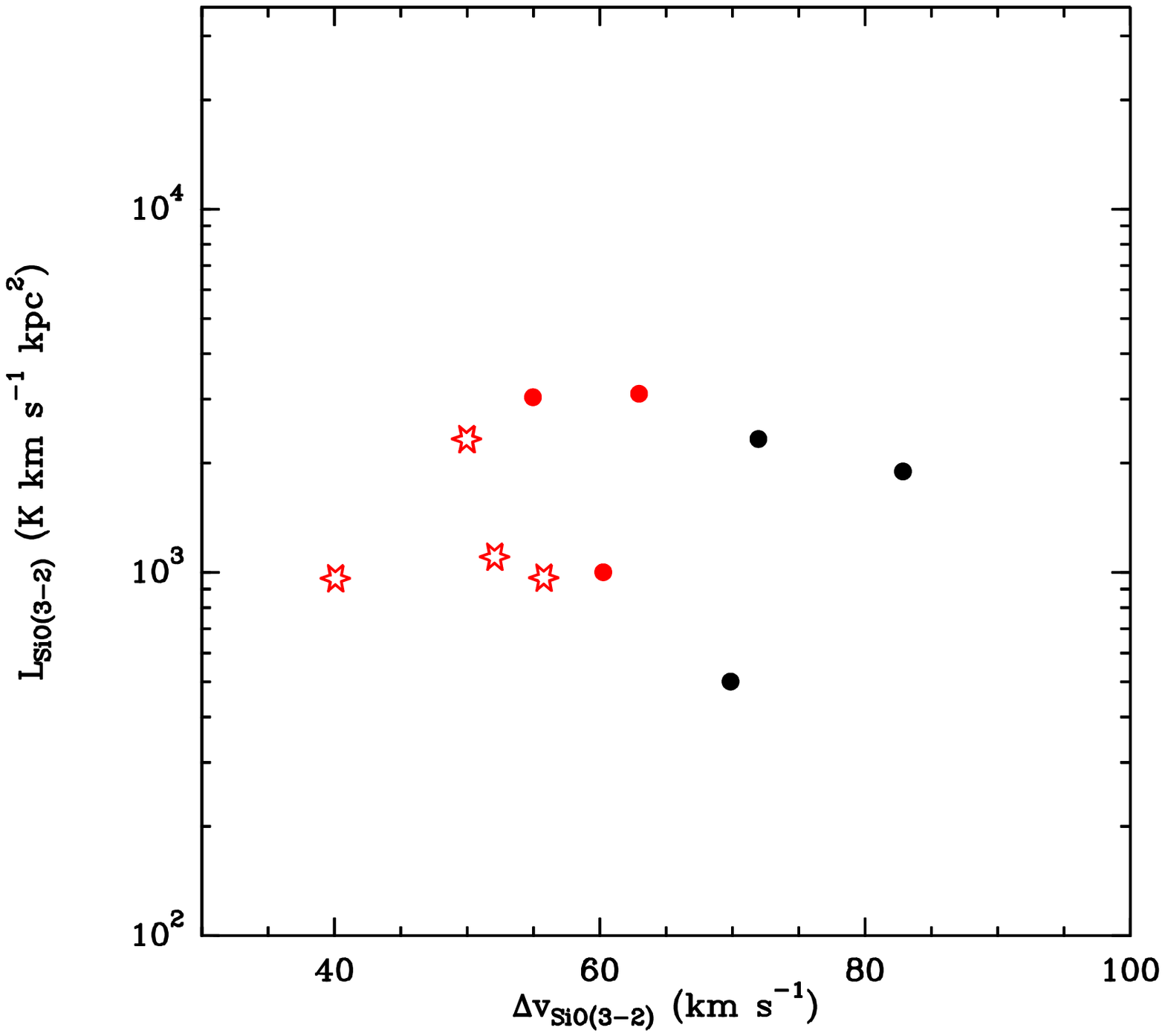}}
\subfigure[][]{\includegraphics[bb = 88 185 535 698,clip,angle=-90,width=6cm]{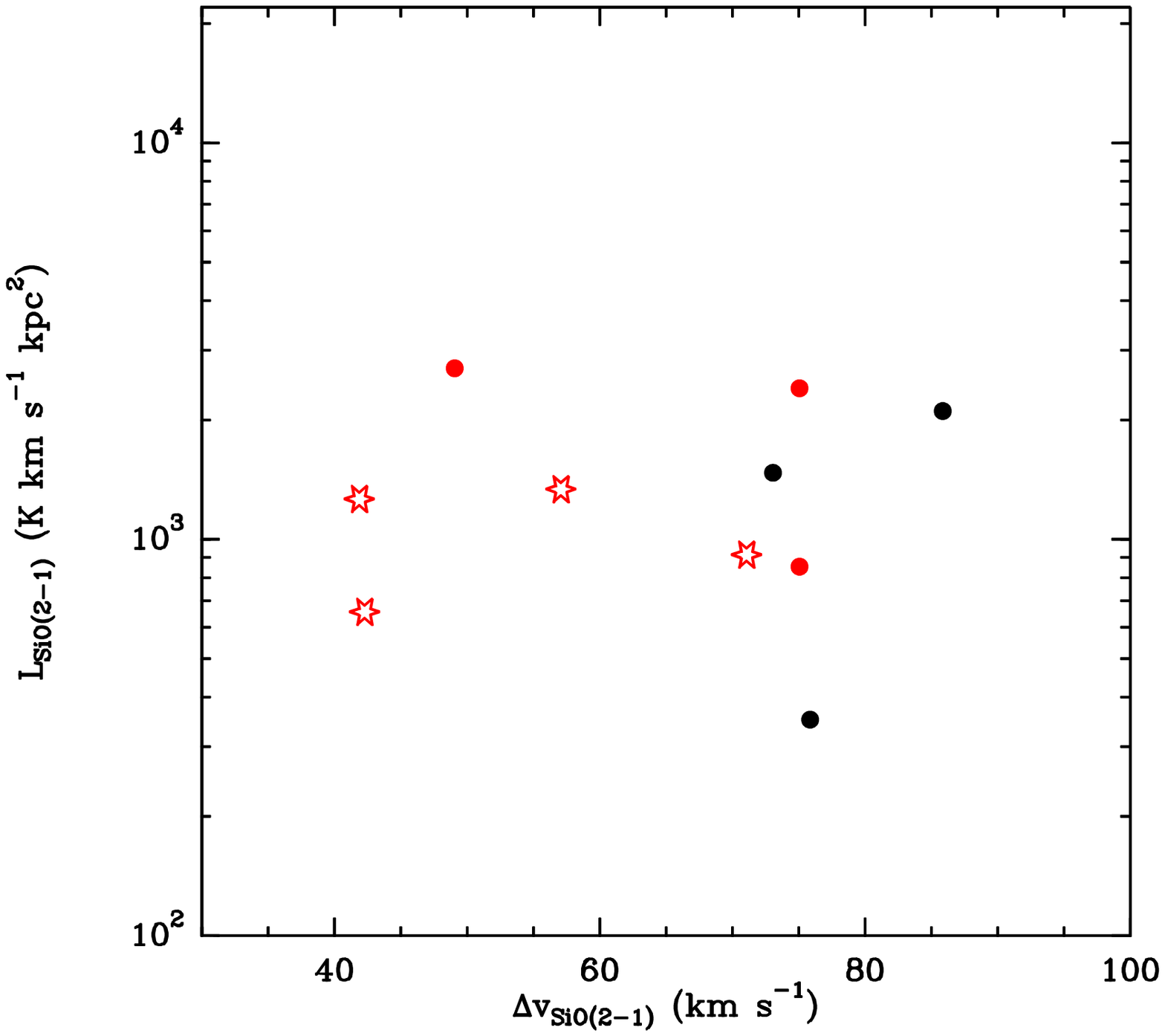}}
\caption{Luminosities of SiO(8--7) ({\bf{a}}), SiO(3--2) ({\bf{b}}) and SiO(2--1) ({\bf{c}}) as a function of their corresponding line width. 22\,$\mu$-quiet and 22\,$\mu$-loud sources are marked as black and red filled circles. Red stars are 8\,$\mu$-loud sources.}\label{dvs}
\end{figure*}

\end{appendix}
\end{document}